\documentclass[twocolumn]{rspublic}
\usepackage[english]{babel}
\usepackage{graphics}
\usepackage{natbib}
\usepackage{amsmath}
\usepackage{amsfonts}
\usepackage{dsfont}
\usepackage{graphicx}
\newcommand{\kvec}[2]{\begin{pmatrix} #1 \\ #2 \end{pmatrix} }
\newcommand{\kvecc}[4]{\begin{pmatrix} #1 \\ #2 \\ #3 \\ #4 \end{pmatrix} }

\newcommand{\matt}[4]{\begin{pmatrix} #1 & #2 \\ #3 & #4 \end{pmatrix} }
\newcommand{\ve}{\varepsilon}
\newcommand{\pdje}[1]{\partial_{#1}}
\begin{document}
%
% title, authors, abstract
%% #AQ1
\title[Review: Single-layer and bilayer graphene SLs]{Single-layer and bilayer graphene superlattices: collimation, additional Dirac points and Dirac lines}
\author[M. Barbier, P. Vasilopoulos, and F. M. Peeters]{Micha\"el Barbier, 
Panagiotis Vasilopoulos, and Fran\c cois M. Peeters}
\affiliation{$^1$Department of Physics, University of Antwerp,\\
Groenenborgerlaan 171, B-2020 Antwerpen, Belgium\\
$^2$Department of Physics, Concordia University,\\ 
7141 Sherbrooke Ouest, Montr\'eal, Quebec, Canada H4B 1R6\\}
\label{firstpage}
\maketitle
\begin{abstract}{graphene; electron transport; two-dimensional crystals}
We review the energy spectrum and transport properties of several types of 
one-dimensional superlattices (SLs) on single-layer and bilayer graphene. 
In single-layer graphene, for certain SL 
parameters an electron beam incident on a SL is highly collimated. On the other 
hand there are extra Dirac points generated for other SL parameters. 
Using rectangular barriers allows us to find analytic expressions for the 
location 
of new Dirac points in the spectrum and for the renormalization of the electron 
velocities. The influence 
of these extra Dirac points on the conductivity is investigated. 
In the limit of $\delta$-function barriers, 
the transmission $T$ through, conductance 
$G$ of a finite number of barriers as well as the energy spectra of SLs are 
{\it periodic} functions of the dimensionless
strength $P$ of the 
barriers, $P \delta(x) = V(x)/ \hbar v_F$, with $v_F$ 
the Fermi velocity. For a Kronig-Penney 
SL with alternating sign of the height of the barriers the 
Dirac point becomes a Dirac line for $P = \pi/2 + n \pi$ with $n$ an integer. 
In bilayer graphene, with an 
appropriate bias applied to the barriers and wells, we show that several new 
types of SLs are produced and two of them are similar to type I and type II 
semiconductor SLs. 
%%R
Similar as in single-layer graphene extra ``Dirac'' points are found.
Non-ballistic transport is also considered.
\end{abstract}

%%%%%%%%%%%%%%%%%%%%%%%%%%%%%%%%%%%%%%%%%%%%%%%%%%%%%%%%%%%%%%%%%%%%%%%%%%%%%
%%%%%
%%%%%
%%%%%%%%%%%%%%%%%%%%%%%%%%%%%%%%%%%%%%%%%%%%%%%%%%%%%%%%%%%%%%%%%%%%%%%%%%%%%
%%%%%
%%%%% Introduction
%%%%%
%%%%%%%%%%%%%%%%%%%%%%%%%%%%%%%%%%%%%%%%%%%%%%%%%%%%%%%%%%%%%%%%%%%%%%%%%%%%%
%%%%%
%%%%%
%%%%%%%%%%%%%%%%%%%%%%%%%%%%%%%%%%%%%%%%%%%%%%%%%%%%%%%%%%%%%%%%%%%%%%%%%%%%%
\section{Introduction}\label{sec0}
%
%GRAPHENE
Since the experimental 
realisation of graphene \citep{geim1} in 2004, 
this one-atom thick layer of carbon atoms has 
attracted the attention of the 
scientific world.
This attraction pole was created by the prediction that the carriers in graphene 
behave as massless relativistic fermions moving in two dimensions. The latter 
particles, which are described by 
the Dirac-Weyl Hamiltonian, possess 
interesting properties 
such as a gapless and linear-in-wave vector electronic spectrum, a perfect 
transmission, at normal incidence, through any potential barrier, i.e., 
the Klein paradox \citep{klein,kat1,miltonreview,gumbs}, which was recently addressed 
experimentally \citep{kex1,kex2}, the zitterbewegung \citep{zit1,zit2,zit3}, 
etc., see Ref.~\citep{cast} 
and \citep{abergel} for  recent reviews. 
On the other hand, in bilayer graphene the carriers 
exhibit a very different but extraordinary electronic behaviour, such as 
being chiral \citep{mccann,kat1} but with a different pseudospin (=1) than in 
single-layer graphene (=1/2). 
Although their spectrum is parabolic in wave vector and also gapless, it is 
possible to create an energy gap by applying a perpendicular electric field on 
a bilayer graphene 
sample \citep{castro}. This allows one to electrostatically create quantum dots 
in bilayer graphene \citep{miltonQdot} and enrich its 
technological capabilities.

In previous work we studied the band structure and other properties of 
single-layer and bilayer graphene \citep{barb1,barb2} in the presence 
of one-dimensional (1D) periodic potential, i.e., a superlattice (SL). 
SLs are known to be useful in altering the band structure of materials and 
thereby broadening their technological applicability.

The already peculiar, cone-shaped band structure of single-layer graphene can 
be drastically changed in a SL. 
An interesting feature is that for 
certain SL parameters the carriers are restricted to move along one direction, 
i.e.\ they are collimated \citep{parksgs}. Furthermore, it was found that for 
other parameters of the SL instead of the single-valley 
( the $K$ or $K'$-point) 
Dirac cone, ``extra Dirac points'' appeared 
at the Fermi level in addition to the 
original one \citep{howiggles}. 
The latter extra Dirac points are interesting because of their accompagning zero 
modes \citep{breywiggles} and 
their influence on many physical properties such as the density 
of states \citep{howiggles}, 
the conductivity \citep{barb4,zhuwiggles}, and the Landau levels upon applying a 
magnetic 
field \citep{parkwiggles, breywiggles3}.

One can also obtain ``extra Dirac points'' in bilayer graphene SLs.
The possibility of locally altering the 
gap \citep{castro} of bilayer graphene by applying a bias 
is another way 
of tuning  the band structure. In this review we classify 
these SLs in four types. 
Another interesting result 
of applying a bias locally is that sign flips of 
the bias introduce bound states along the interfaces \citep{mart, jalil}. 
These bound states break the time reversal 
symmetry and are distinct for the two $K$ and $K'$ valleys; 
this opens up perspectives for valley-filter devices \citep{sanjose}. 

In this review we will use the following methods to describe our 
findings.
For both single-layer and bilayer graphene we will use the 
nearest neighbour, tight-binding Hamiltonian in the continuum approximation, 
and restrict ourselves to the electronic 
structure in the neighbourhood of the $K$ point. We then apply the 
transfer-matrix method
to study the spectrum of and transmission through various potential barrier 
structures, which we approximate by
piecewise constant potentials. 
We consider structures with a finite number of barriers and SLs.

We will study ballistic transport in systems with a finite number of barriers 
using 
the two-probe Landauer conductance while in a SL (infinite number of 
barriers) we will evaluate the spectrum 
and the diffusive conductivity, i.e., we will study 
non-ballistic transport. 

The work is organized as follows. In Sec.~\ref{sec1} we investigate 
various aspects of ballistic transport through a finite number of barriers on 
single-layer 
graphene as well as the spectrum of SLs, with emphasis on collimation and extra 
Dirac points and 
their influence on 
non-ballistic transport. In Sec.~\ref{sec2} we  carry on the 
same studies, whenever possible, on bilayer graphene. In addition, we consider 
various types of band alignments in the presence of a bias that can lead to 
different types of heterostructures and SLs. 
We make a summary and concluding remarks in Sec.~\ref{sec3}.

%%%%%%%%%%%%%%%%%%%%%%%%%%%%%%%%%%%%%%%%%%%%%%%%%%%%%%%%%%%%%%%%%%%%%%%%%%%%%
%%%%%
%%%%%
%%%%%%%%%%%%%%%%%%%%%%%%%%%%%%%%%%%%%%%%%%%%%%%%%%%%%%%%%%%%%%%%%%%%%%%%%%%%%
%%%%%
%%%%% SECTION 1
%%%%%
%%%%%%%%%%%%%%%%%%%%%%%%%%%%%%%%%%%%%%%%%%%%%%%%%%%%%%%%%%%%%%%%%%%%%%%%%%%%%
%%%%%
%%%%%
%%%%%%%%%%%%%%%%%%%%%%%%%%%%%%%%%%%%%%%%%%%%%%%%%%%%%%%%%%%%%%%%%%%%%%%%%%%%%
\section{Single-layer graphene}\label{sec1}

We describe the electronic structure of an infinitely large, flat graphene flake 
by the nearest-neighbour tight-binding model 
and consider 
wave vectors close to the K point. 
The relevant Hamiltonian in the continuum approximation is 
$\mathcal{H} = v_F {\boldsymbol\sigma} \cdot {\bf \hat{p}} + 
V \mathds{1} + m v_F^2 \sigma_z$, with 
${\bf \hat{p}}$ the momentum operator, 
$V$ the potential, $\mathds{1}$ the $2\times 2$ unit matrix, 
${\boldsymbol\sigma} = (\sigma_x \sigma_y)$, $\sigma_z$ the Pauli-matrices
and $v_F \approx 10^6 m/s$ the Fermi velocity. Explicitly $\mathcal{H}$ is 
given by
\begin{equation}\label{eq1_1}
	\mathcal{H} = \matt{V + m v_F^2}{-\ri v_F \hbar (\pdje{x}-\ri \pdje{y})}{-\ri v_F \hbar (\pdje{x}+\ri \pdje{y})}{V - m v_F^2}.
\end{equation}
The mass term is in principle zero in the nearest-neighbour, tight-binding model 
but due to interaction  with a substrate \citep{gio} an effective mass term can be induced and 
results in the opening of an energy gap. Recently there have been proposals 
to induce an energy gap in single-layer graphene, and 
it is appropriate that we consider this mass term where relevant.
In the presence of a 1D rectangular potential $V(x)$, such  as 
the one shown in Fig. \ref{fig1_1}, the equation $(\mathcal{H} - E)\psi = 0$ 
admits (right- and left-travelling) plane wave solutions of the form 
$\psi_{l,r}(x) e^{i k_y y}$ 
with
\begin{equation}\label{eq1_2}
	\psi_r(x) = \kvec{\ve + \mu}{\lambda + \ri k_y}\re^{\ri \lambda x}, \,  \psi_l(x) = \kvec{\ve + \mu}{-\lambda + \ri k_y} \re^{-\ri \lambda x},
\end{equation}
here 
$\lambda = [(\ve-u(x))^2 - k_y^2 - \mu^2]^{1/2}$ is the $x$ component of the wave vector, 
$\ve = EL/\hbar v_F$, $u(x) = V(x)L/\hbar v_F $, and $\mu = m v_F L/\hbar$. 
The dimensionless parameters $\ve$, $u(x)$ and $\mu$ scale with the 
characteristic length $L$ of the potential barrier structure. 
For the single or  double barrier system this $L$ will be equal to the barrier 
width while for 
a SL it will be its period. 
Neglecting the mass term one rewrites Eq. (\ref{eq1_2}) in the simpler form 
\begin{equation}\label{eq1_3}
	\psi_r(x) = \kvec{1}{s \re^{\ri \phi}}\re^{\ri \lambda x}, \quad  \psi_l(x) = \kvec{1}{-s \re^{-\ri \phi}}\re^{-\ri \lambda x},
\end{equation}
with $\lambda = [(\ve-u(x))^2 - k_y^2]^{1/2}$,\,\,$\tan \phi = k_y/\lambda$, and  
$s = \sgn(\ve-u(x))$.

% sec1a
%%%%%%%%%%%%%%%%%%%%%%%%%%%%%%%%%%%%%%%%%%%%%%%%%%%%%%%%%%%%%%%%%%%%%%%%%
\subsection{A single or double barrier}\label{sec1a}
\begin{figure}[ht]
  \begin{center}
  	\includegraphics[width=8cm]{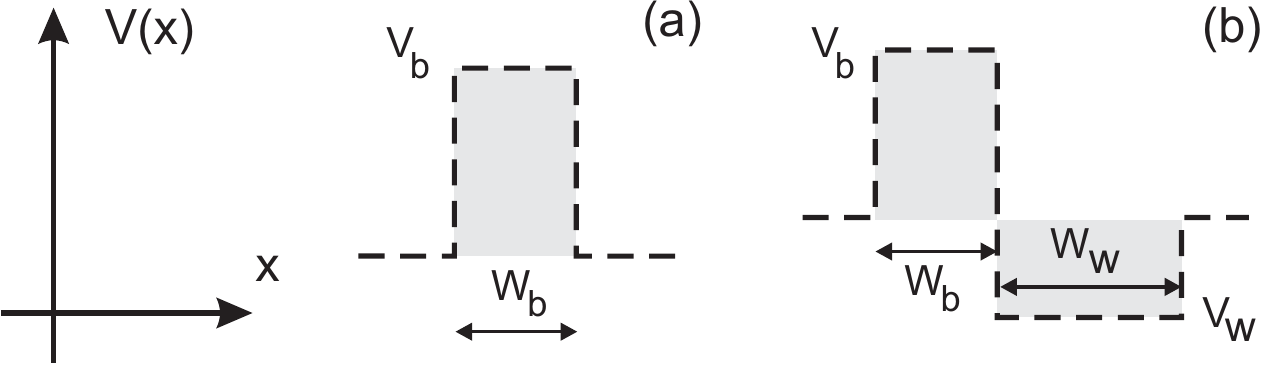}
  \end{center}
  \caption{(a) A 1D potential barrier of height $V_b$ and width $W_b$. (b) A single unit of a potential well next to a potential barrier. 
    }\label{fig1_1}
\end{figure}
The model barriers and wells we consider 
are shown in Fig. \ref{fig1_1}(a). It is 
interesting to look at the tunneling through such barriers, which was previously 
studied by \citep{kat1} for a single barrier. 
This was later extended to 
massive electrons with spatially varying mass \citep{gomesmass}.

%%R
% Following the approach of 
% \citep{breywiggles2} we can rewrite 
% the time-independent Schr\"odinger equation $\mathcal{H}\psi = \ve \psi$ 
% as $\pdje{x}\psi(x) = C(x)\psi(x)$, with $C(x)$ the coefficient matrix
% \begin{equation}\label{eq1a_-1}
% 	C(x) = \matt{-k_y}{\ri(\ve-u(x)-\mu)}{\ri(\ve-u(x)+\mu)}{k_y}.
% \end{equation}
% This equation has the solution
% \begin{equation}\label{eq1a_0}
% 	\psi(x) = \mathcal{P} \exp\{ \int^x_0 \rd x' C(x') \}\psi(0),
% \end{equation}
% where $\mathcal{P}$ is the 
% path-ordering operator. 
% For a piecewise constant potential $u(x) = u$ this factor becomes a product of the 
% characteristic matrices of the various regions of constant potential. 

{\it Transmission.} 
%%%%%%%%%%%%%%%%%%%%%%%%%%%%%%%%%%%%%%%%%%%%%%%%%%%%%%%%%%%%%%%%%%%%%%%%%%%%%%
%
To find the transmission $T$ through a square-barrier 
structure one first observes that 
the wave function in the $j$th region $\psi_j(x)$ of the constant potential $V_j$ 
is given by a superposition of the eigenstates given by Eq.~(\ref{eq1_2}),
\begin{equation}\label{eq1a_1}
	\psi_j(x) = A_j {\psi_r}_{j} + B_j {\psi_l}_{j}.
\end{equation}
The wave function should be continuous at the interfaces.
This boundary condition gives the transfer matrix 
$\mathcal{N}_j$ relating the coefficients $A_j$ and $B_j$ of region $j$ with 
those of the region $j+1$ in the manner 
\begin{equation}\label{eq1a_2}
	\kvec{A_j}{B_j} = \mathcal{N}_{j+1} \kvec{A_{j+1}}{B_{j+1}}.
\end{equation}
By employing the transfer matrix at each potential step we obtain, after $n$ steps, the relation
\begin{equation}\label{eq1a_3}
	\kvec{A_0}{B_0} = \prod^n_{j = 1} \mathcal{N}_{j} \kvec{A_{n}}{B_{n}}.
\end{equation}
In the region to the left of the barrier we assume $A_0 = 1$ and 
denote by $B_0 = r$ the reflection amplitude.  Likewise, to the right of the 
$n$th barrier we have $B_n = 0$ and denote by $A_n = t$ the transmission 
amplitude.

\begin{figure}[ht]
  \begin{center}
  	\includegraphics[width=11cm]{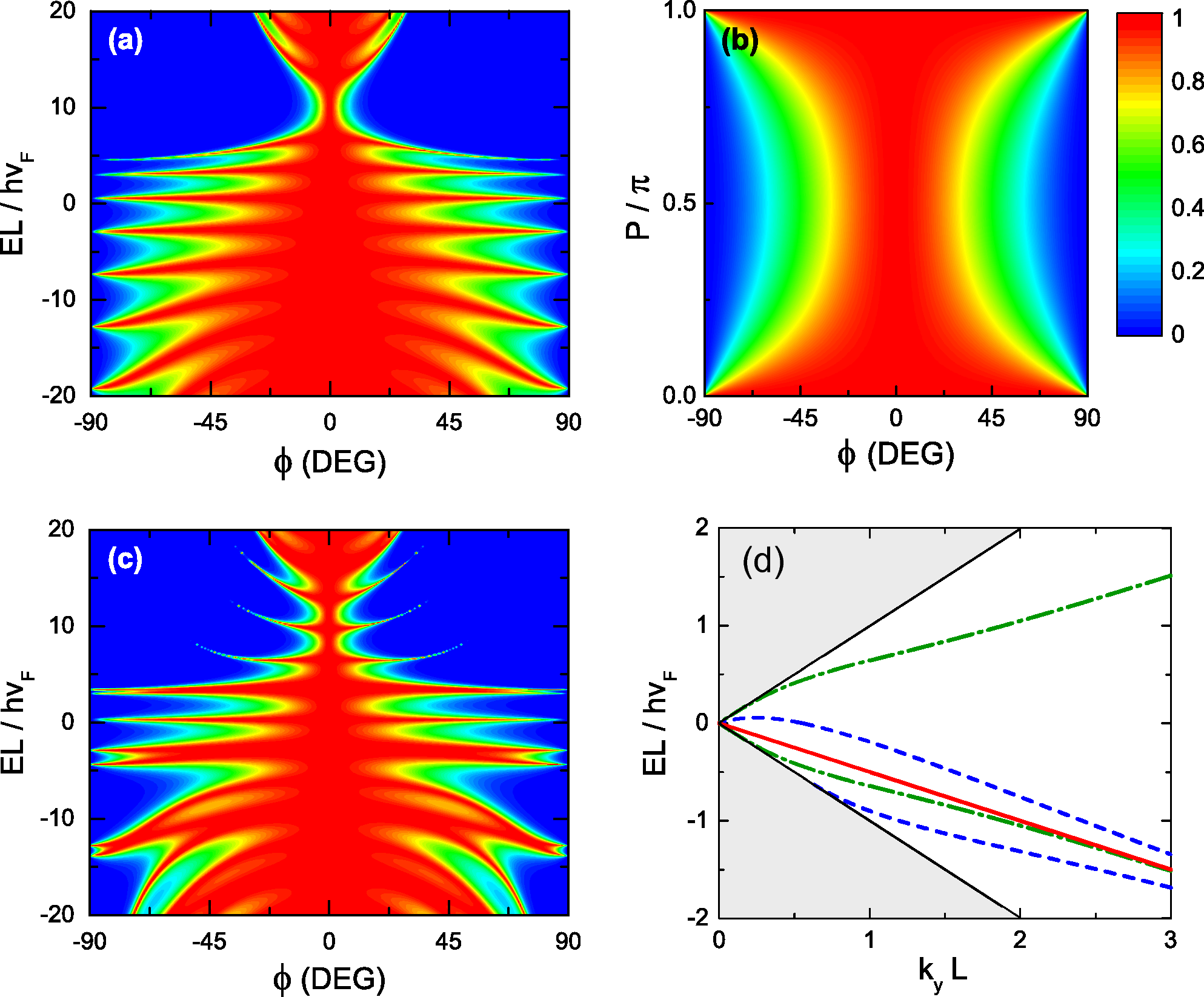}
  \end{center}
  \caption{(a) Contour plot of the transmission through a single barrier with 
$\mu = 0$, $W_b = L$, and $u_b = 10$. (b) As in (a) for a single 
$\delta$-function  barrier with $\mu = 0$ and $u(x) = P \delta(x)$; 
the  transmission is independent of the energy. (c) As in (a) for two barriers 
with $\mu = 0$, $u_b = 10$, $u_w = 0$, $W_b = 0.5 L$, and $W_w = L$. 
(d) Spectrum of the bound states vs $k_y$ for a single 
($L = 1$, solid red line), two parallel (dashed blue curves), 
and two anti-parallel (green dash-dotted curves) $\delta$-function 
barriers ($L$ is the inter-barrier distance).}\label{fig1_2}
\end{figure}
% %
The transmission probability $T$ can be expressed as 
the ratio of the transmitted current density $j_x$ over the incident one, 
where $j_x = v_F \psi^\dagger \sigma_x \psi$. This results 
in $T = (\lambda'/\lambda)|t|^2$, with $\lambda'/\lambda$ the ratio between the 
wave vector 
%%R
$\lambda'$ to the right and 
%%R
$\lambda$ to the left 
of the barrier. 
If the potential to the right and left of the barrier 
is the same we have 
$\lambda' = \lambda$. 
For a single barrier the transmission amplitude is given by 
$T = |t|^2 = |N_{11}|^{-1}$, 
with $N_{ij}$ the elements of the transfer matrix $\mathcal{N}$. 
Explicitly, $t$  can be written as 
\begin{equation}\label{eq1a_4}
\begin{aligned}
	& 1/t = \cos(\lambda_b W_b) - i Q \sin(\lambda_b W_b),\\
	& Q = (\ve_0 \ve_b - k_y^2 -\mu_0 \mu_b)/\lambda_0 \lambda_b;
\end{aligned}
\end{equation}
the indices $0$ and $b$ refer, respectively, to the region outside and 
inside the barrier and $\ve_b = \ve - u$. 
A contour plot of the transmission is 
shown in Fig.~\ref{fig1_2}(a). 
We clearly see: 1) $T = 1$ for $\phi = 0$ 
which is the well-known Klein tunneling, and 2) strong resonances, in particular 
for $E <0$, 
%%R
when $\lambda_b W_b = n \pi$, 
which describe 
hole scattering above a potential well.

In the limit of a very thin and high barrier, one can model it 
by a $\delta$-function barrier $V(x) /\hbar v_F = P \delta(x)$. 
Using Eq.~(\ref{eq1a_4}) for $t$ gives \citep{barb3}
\begin{equation}\label{eq1a_5}
    T = 1/[1 + \sin^2 P \tan^2 \phi],
\end{equation}
with $\tan \phi = k_y/\lambda_0$ the angle of incidence. 
Notice that this transmission is independent of the energy and is a 
{\it periodic} function of $P$. The latter is very different from the 
non-relativistic case where T is a decreasing function of P. A contour plot of 
the transmission is shown in Fig.~\ref{fig1_2}(b) and $T = 1$ for $\phi \approx 0$  
which is nothing else than Klein tunneling. Notice also the symmetry $T(\pi - P) = T(P)$.

For two barriers the system becomes a resonant structure, for which it was found 
that the resonances in the transmission depend mostly on the width $W_w$ of 
the well between the barriers \citep{miltondouble}. A plot of the transmission 
is shown in Fig.~\ref{fig1_2}(c). In the limit of two parallel $\delta$-function 
barriers of equal strength $P$ we obtain  the transmission
\begin{equation}\label{eq1a_6}
    T = \big[1+ \tan^2\phi ( \cos \lambda_0 \sin 2 P - 2 s \sin \lambda_0 \sin^2 P 
    /\cos\phi)^2 \big]^{-1}.
\end{equation}
The case of two anti-parallel $\delta$-function barriers of equal strength is 
also interesting. The relevant transmission is 
\begin{equation}\label{eq1a_7}
	T = \big[ \cos^2 \lambda_0 + \sin^2 \lambda_0 (1 - \sin^2 \phi \cos 2 P)^{2}/
	\cos^4 \phi \big]^{-1}.
\end{equation}

\begin{figure}[ht]
	\begin{center}
	  	\includegraphics[width=8cm]{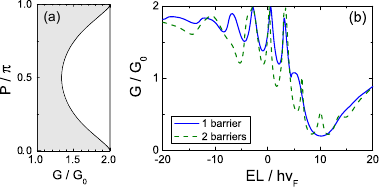}
	\end{center}
	\caption{(a) Conductance $G$ vs strength $P$ of a  $\delta$-function 
barrier in  single-layer graphene; 
	the conductance is independent of the energy. (b) Conductance $G$ vs 
energy for the single (solid blue  curve) and double (dashed green curve) 
square barrier of Fig.~\ref{fig1_2}(a) and \ref{fig1_2}(c).}
	\label{fig1_3}
\end{figure}

{\it Conductance.} The two-terminal conductance is given by
\begin{equation}\label{eq1a_8}
    G(E_F) = G_0\int_{-\pi/2}^{\pi/2} T(E_F,\phi) \cos\phi\, \rd \phi,
\end{equation}
with $G_0 = 2 E_F L_y e^2/ (v_F h^2)$ for single-layer graphene, and $L_y$ the 
width of the system.  For a single and double barrier, the transmission through 
which is plotted 
in Fig.~\ref{fig1_2}(a) and \ref{fig1_2}(c), the conductance $G$ is shown in 
Fig.~\ref{fig1_3}(b) and exhibits multiple resonances despite the integration 
over the angle $\phi$. 

Taking the limit of a $\delta$-function barrier leads to $G$ periodic in $P$ and given by
\begin{equation}\label{eq1a_9}
    G/G_0 = 2\big[ 1 - \text{artanh}(\cos P) \sin P \tan P\big]/\cos^2 P.
\end{equation}
For one period $G$ is shown in Fig.~\ref{fig1_3}(a).

\begin{figure}[ht]
  \begin{center}
	\includegraphics[width=11.0cm]{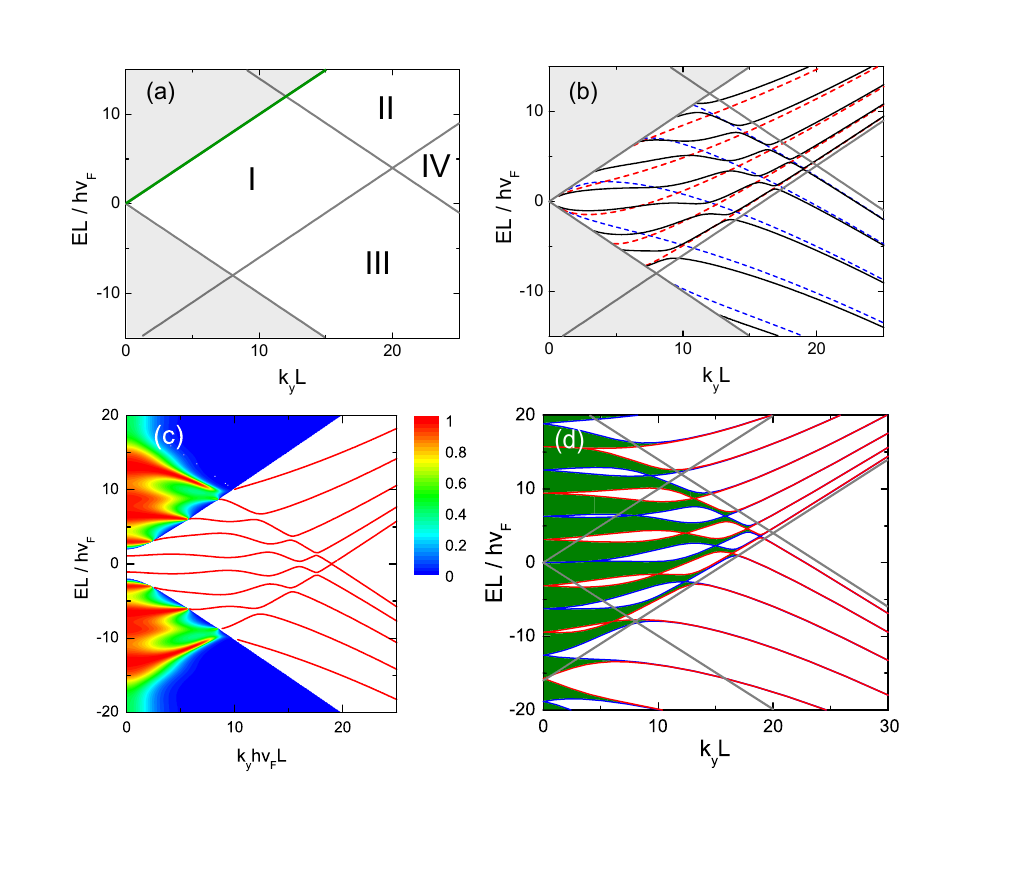}
  \end{center}
  \caption{
  (a) Four different regions for a single unit of Fig. \ref{fig1_1}(b) with 
$u_b = 24$, $u_w = 16$, 
$W_b = 0.4$ and $W_w = 0.6$. 
The green line corresponds to region I in the limit of a 
$\delta$-function barrier. 
  (b) 
  Bound states for a single 
   barrier (dashed blue curves) and well (dashed red curves) and the combined 
barrier-well unit (black curves). 
  (c) Contour plot of the transmission through a 
  unit with $\mu = 2$, $u_b = - u_w = 20$ and 
  $W_b = W_w = 0.5$; the red curves show 
  the bound states. 
  (d) Spectrum of a SL whose 
   unit cell is shown in Fig. 1(b), 
   for $k_x = 0$ (blue curves) and 
   %%R
   $k_x L = \pi/2$ (red curves).
  }\label{fig1_4}
\end{figure}

{\it Bound states.}  For $k_y^2 + \mu_0^2 > \ve^2$ the wave function outside the 
barrier (well) becomes an exponentially decaying function of $x$, 
$\psi(x) \propto \exp\{\pm|k_x| x\}$ with $|k_x| = [k_y^2 + \mu_0^2 - \ve^2]^{1/2}$. 
Localized states form near the barrier boundaries \citep{miltonwell}; however, 
they are propagating freely along the $y$-direction. The 
spectrum of these bound states can be found by 
setting the determinant of the transfer matrix equal to zero. For a 
single potential barrier (well) it is given by the solution of the  transcendental equation
\begin{equation}
	|\lambda_0| \lambda_b \cos(\lambda_b W_b)+(k_y^2 + \mu_0 \mu_b -\ve(\ve-u)) \sin(\lambda_b W_b) = 0.
\end{equation}
In Fig.~\ref{fig1_4}(b) these bound states are shown, 
as a function of $k_y$, by the dashed blue (red)  curves. 

An interesting structure to study is that of a potential barrier next to a 
well but with average potential equal to zero, 
%%R
considered by 
%%R and studied in greater detail by 
\citep{breywiggles2}. This is the unit cell (shown in Fig.~\ref{fig1_1}(b)) 
of the SL 
we will use in Sec.~\ref{sec2} where extra Dirac points will be found. 
In Fig.~\ref{fig1_4}(a) the Dirac cone outside the barrier is shown as a grey 
area, inside this region there are no bound states. 
Superimposed are grey lines 
corresponding to the edges of the Dirac cones inside the well and barrier 
that divide the $(E,k_y)$ plane into four regions. Region I corresponds to 
propagating states inside both the barrier and well while region II (III) 
corresponds 
to  propagating states only inside the well (barrier). 
In region IV no propagating 
modes are possible, neither in the barrier nor in the well. For high thin 
barriers, region I will become a thin area adjacent to the upper cone, 
converging to the dark green line in the limit of a $\delta$-function barrier.
Figure~\ref{fig1_4}(b) shows that the bound states of this structure are 
composed of 
the ones of a single barrier and those 
of a single well. Anticrossings take 
place where the bands otherwise would cross. 
The resulting spectrum is clearly a starter of the 
spectrum of a SL shown in Fig.~\ref{fig1_4}(d).

In the limit of $\delta$-function barriers and wells 
the expressions for the dispersion relation are strongly simplified
by setting $\mu = 0$ in all regions.
For a single $\delta$-function barrier the bound state is given by
\begin{equation}
	\ve = \sgn(\sin P) |k_y| \cos P,
\end{equation}
which is a straight line with a reduced group velocity $v_y$; 
the result is shown in Fig.~\ref{fig1_2}(d) by the red curve. 
Comparing with the single-barrier case we notice that due to the 
periodicity in $P$,  the $\delta$-function barrier can act as a barrier or as a 
well depending on the value of $P$.  

For two $\delta$-function barriers there are two important cases: 
the parallel and the anti-parallel case. For parallel barriers one finds an implicit 
equation for the energy
\begin{equation}
	|\lambda' \cos P + \ve \sin P| = |\re^{-\lambda'%%R L
	} k_y \sin P|,
\end{equation}
where $\lambda' = |\lambda_0|$, 
while for anti-parallel barriers one obtains
\begin{equation}
	k_y^2 \sin^2 P=\lambda'^2/(1 - \re^{-2 \lambda'%%R L
	}).
\end{equation}
For two (anti-)parallel $\delta$-function barriers we have, for each fixed $k_y$ 
and $P$, two energy values $\pm\ve$, and therefore 
two bound states. 
In both cases, for $P = n\pi$ 
the spectrum is simplified 
to the one in the absence of any potential 
$\ve = \pm |k_y|$. 
In Fig.~\ref{fig1_2}(d) the bound states for double 
(anti-)parallel $\delta$-function barriers 
are shown, as a function of $k_y L$, by the dashed blue (dash-dotted green) 
curves. 
For anti-parallel barriers we 
see that there is a symmetry around $E = 0$, which is absent when the  barriers 
are parallel.
%
% sec1b
%%%%%%%%%%%%%%%%%%%%%%%%%%%%%%%%%%%%%%%%%%%%%%%%%%%%%%%%%%%%%%%%%%%%%%%%%
\subsection{Superlattice}\label{sec1b}
%%R
% For a number of units $N$ the tranfer matrix can be calculated by using 
% Chebyshev polynomials \citep{chebyshevmeth1,chebyshevmeth2}; it is given by the 
% $N$th power of the characteristic (unimodular) matrix $\mathcal{Q}$ (with matrix 
% elements $q_{ij}$) as
% %
% \begin{equation}\label{eq1b_1}
% 	\mathcal{Q}^N = \matt{q_{11} U_{N-1}(a)-U_{N-2}(a)}{q_{12} U_{N-1}(a)}{q_{21} U_{N-1}(a)}{q_{22}U_{N-1}(a) -U_{N-2}(a)},
% \end{equation}
% %
% where $a = 1/2(q_{11} + q_{22})$; $U_{m}(x)$ are the 
% Chebyshev polynomials of the second kind given by 
% $U_{m} (x)= \sin[(m+1)\gamma]/\sin \gamma$ with $\gamma = \arccos x$. 
% We notice that for propagating states outside the barriers, i.e., for
% $\ve^2 > k_y^2 + \mu^2$, we have  $q_{22} = q_{11}^*$ and 
% $q_{21} = -q_{12}^*$. Then  the transmission can be calculated as before 
% and takes the form \citep{griffiths}
% %
% \begin{equation}\label{eq1b_2}
% 	T = \frac{1}{1 + |q_{12}|^2 U^2_{n-1}(a)},
% \end{equation}
% with $a = \Real\{q_{11} \re^{-\ri \lambda_0 L}\}$. 
%
%%R
% In the case of a SL we 
% take $N \rightarrow \infty$. 
% The corresponding 1D 
Now we will consider the system of a superlattice with a corresponding 1D 
periodic potential, with square barriers, 
%%R
% is 
given by
\begin{equation}\label{eq1b_3}
	V(x) = V_0 \sum_{j = - \infty}^\infty [\Theta(x - j L) - \Theta(x - j L - W_b)].
\end{equation}
with $\Theta(x )$ the step function. The corresponding wave function 
is a Bloch function and satisfies the periodicity condition 
$\psi(L) = \psi(0) \exp(i k_x)$, with  $k_x$ now the Bloch phase. 
Using this relation together with the transfer matrix for a single unit 
$\psi(L) = \mathcal{M}\psi(0)$ leads to the condition
\begin{equation}\label{eq1b_4}
	\det[\mathcal{M} - \exp(i k_x)] = 0.
\end{equation}
This gives  the 
transcendental equation
\begin{equation}\label{eq1b_5}
	\cos k_x = \cos\lambda_w W_w \cos\lambda_b W_b - Q \sin\lambda_w W_w \sin\lambda_b W_b,
\end{equation}
from which we obtain 
the energy spectrum of the system. 
In Eq.~(\ref{eq1b_5}) we used the following notation:
\begin{eqnarray*}\label{eq1b_6}
\nonumber
%%R
	\ve_w = \ve + u W_b, \quad \ve_b = \ve - u W_w, \quad u = V_0 L /\hbar v_F, \quad W_{b,w} \rightarrow W_{b,w}/L,\\
	\lambda_w = [\ve_w^2 - k_y^2 - \mu_w^2]^{1/2}, \quad \lambda_b = [\ve_b^2 - k_y^2 - \mu_b^2]^{1/2}, \quad	Q = (\ve_w \ve_b - k_y^2 - \mu_b\mu_w)/\lambda_w \lambda_b.
\end{eqnarray*}

Numerical results for the 
dispersion relation $E (k_y)$ are shown in 
Fig.~\ref{fig1_4}(d). We see the appearance of bands (green areas) which 
for large $k_y$ values collapse into the bound states (where the red and blue 
curves meet) 
while the charge carriers move freely along 
the $y$ direction.
%
% sec1c
%%%%%%%%%%%%%%%%%%%%%%%%%%%%%%%%%%%%%%%%%%%%%%%%%%%%%%%%%%%%%%%%%%%%%%%%%
\subsection{Collimation and extra Dirac points}\label{sec1c}
As shown by various studies, carriers in graphene SLs
exhibit several interesting pecularities that 
result from the particular electronic SL band structure. 
In a 1D SL it was found that the spectrum can be altered anisotropically 
%%R
\citep{parkanisotropy, noriwiggles}. 
Moreover, this anisotropy 
can be made very large such that 
for a broad region in {\bf k} space the spectrum is 
dispersionless in one direction, and thus electrons are collimated along the 
other direction \citep{parksgs}. 
Even more intriguing was the ability to split off ''extra Dirac points'' 
\citep{howiggles} with accompanying zero modes \citep{breywiggles} which move away 
from the K point along the 
$k_y$ direction with increasing potential strength. Here we will describe these 
phenomena for a SL 
of square potential barriers. 

We start by describing the collimation as done 
by \citep{parksgs}; 
subsequently we will find the conditions on the
parameters of the SL  for which a collimation appears. It turns out that they 
are the same as those needed to create 
a pair of extra Dirac points. 

\begin{figure}[ht]
  \begin{center}
	\includegraphics[width=10cm]{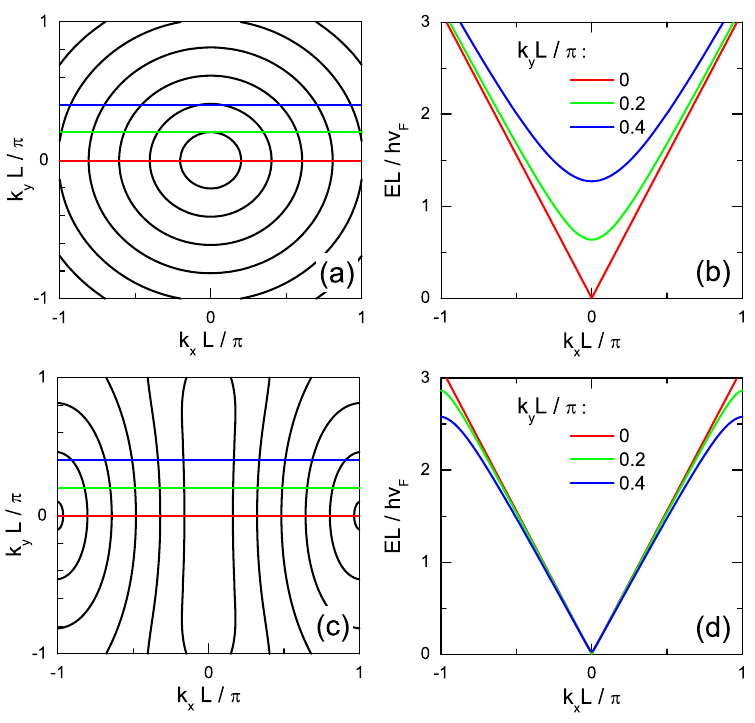}
  \end{center}
  \caption{The lowest conduction band of the spectrum of graphene near the 
  K point in the absence of SL
  potential  (a), (b) and in its presence 
  (c), (d) with 
  %%R
  $u = 4 \pi$. (a) and (c) are contour plots of the conduction band with a contour step of $0.5$ $\hbar v_F / L$. 
  (b) and (d) show slices along constant 
  %%R
  $k_y L = 0,  0.2, 0.4 \pi$.
  }\label{fig1_5}
\end{figure}
Following \citep{parksgs} the condition for collimation to occur is 
$\int_{BZ} \re^{\ri s \hat{s}\alpha(x)} = 0$, 
where the function $\alpha(x) = 2\int^x_0 u(x') \rd{x'}$ embodies
the influence of the potential, $s = \text{sign}(\ve)$ and $\hat{s} = \text{sign}(k_x)$.
%%R
%The condition for this to occur is   
%$\int_{BZ} \re^{\ri s \hat{s}\alpha(x)} = 0$. 
For a symmetric rectangular lattice this corresponds to $u /4 = n \pi$. 
The spectrum for the lowest energy bands is then given by \citep{parknewgen}
\begin{equation}\label{eq1c_4}
	\ve \approx \pm  
	\big[k_x^2 + |f_l|^2 k_y^2 \big]^{1/2} + \pi l / L
	%%R	G_0/2.
\end{equation}
with $f_l$ being the coefficients of the Fourier expansion 
$\re^{\ri\alpha(x)} = \sum_{l = -\infty}^\infty f_l \re^{\ri 
%%R G_0
2 \pi l x / L}$. The 
coefficients 
$f_l$ depend on the potential profile $V(x)$, with $|f_l| < 1$. 
For a symmetric SL of square barriers we have 
%%R
$f_l = u \sin(l\pi/2 - u/2)/(l^2 u^2 - u^2/4)$. 
The inequality $|f_l| < 1$ implies 
a group velocity in the $y$ direction $v_y < v_F$ which can be seen from Eq.~(\ref{eq1c_4}). 

In Fig. \ref{fig1_5}(b),(d) we show the dispersion relation $E$ vs $k_x$  for 
%%R
$u = 4 \pi$ 
at constant $k_y$. As can be seen, when a SL is present in most 
of the Brillouin zone the spectrum, partially shown in (c), is nearly 
independent of 
$k_y$. That is, we have collimation of an electron beam along the SL axis. 
The condition 
$u = V_0 L /\hbar v_F = 4 n \pi$ shows 
that altering the period of the SL \emph{or} the potential height of the 
barriers is sufficient to 
produce collimation. 
This makes a SL a versatile tool for tuning the spectrum.
Comparing with Figs. \ref{fig1_5}(a), (b) we see that the cone-shaped spectrum 
for $u = 0$, is transformed into a wedge-shaped spectrum \citep{parksgs}.

\begin{figure}[ht]
  \begin{center}
	\includegraphics[width=10cm]{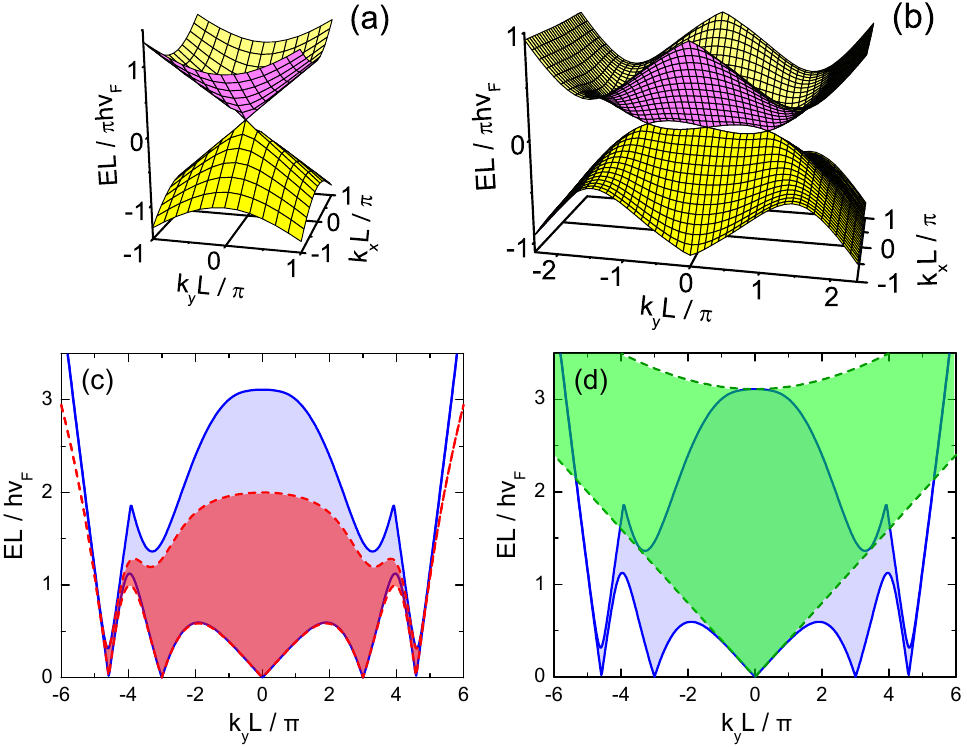}
  \end{center}
  \caption{The spectrum of graphene near 
  		the K point in the absence of a 
  		SL  (a) and in its presence (b) with 
  		%%R
  		$u = 4.5 \pi$. 
  		(c) and (d) The 
  		SL spectrum with 
  		%%R
  		$u = 10 \pi$, 
  		the lowest conduction bands are coloured in cyan, 
 		red, and green for, respectively, the exact, and the approximations 
 		given by (c) Eq.~(\ref{eq1c_6}) and (d) Eq.~(\ref{eq1c_7}), 
 		respectively. The approximate spectra are delimited by the dashed 
 		curves.
  }\label{fig1_6}
\end{figure}

We will compare this result now with an 
%%R
other approximate result for the spectrum, 
where we suppose $\ve$ small instead of $k_y$ small. 
We start with the transcendental equation (\ref{eq1b_5}). 
As we are interested in an analytical approximate expression for the spectrum, 
we choose to expand the dispersion 
relation around $\ve = 0$ up to second order in $\ve$. The resulting spectrum is 
\begin{equation}\label{eq1c_6}
	\ve_\pm = \pm 
	\left[{\frac {4|a^2|^2\,\big[k_y^2 \sin^2 (a/2) + a^2 \sin^2(k_x/2)\big]}{k_y^4 a \sin a + a^2 u^4/16 - 2 k_y^2 u^2 \sin^2 (a/2) 
	}}\right]^{1/2},
\end{equation}
with $a = [u^2/4 - k_y^2]^{1/2}$. 
In order to compare this  spectrum with that by 
\citep{parksgs}, we 
expand Eq.~(\ref{eq1b_5}) for small ${\bf k}$ and $\ve$; 
this leads to
\begin{equation}\label{eq1c_7}
	\ve \approx \pm \big[k_x^2 + k_y^2\, \sin^2(u/4)/(u/4)^2\big]^{1/2}.
\end{equation}
This spectrum has the form of an anisotropic cone and 
corresponds to that of Eq.~(\ref{eq1c_4}) 
for $l = 0$ (higher $l$ correspond to higher energy bands). 
In Fig.~\ref{fig1_6}(a), (b) we see that the cone-shaped spectrum in (a), for 
$u = 0$, is transformed into a  anisotropic spectrum in (b), 
for 
%%R
$u = 4.5 \pi$, 
having peculiar extra Dirac points. These extra Dirac points 
cannot be described by a spectrum having an anisotropic cone-shape, therefore we 
compare the two approximate spectra. 
In Fig.~\ref{fig1_6}(c), (d) we 
show how 
Eq.~(\ref{eq1c_6})) and 
Eq.~(\ref{eq1c_7}) differ from the ``exact'' numerically obtained spectrum. 
From this figure one can 
see that Eq.~(\ref{eq1c_6}) describes the lowest bands rather well for 
$\ve < 1$, while Eq.~(\ref{eq1c_7}) is sufficient to describe the spectrum near 
the Dirac point. 
The former equation will be usefull when describing the 
spectrum near the extra Dirac points and we will use it 
to obtain the velocity.

We  now move on to another important feature of the spectrum, 
the extra Dirac points first obtained by \citep{howiggles}
%%&Ho \emph{et al.} 
using tight-binding calculations. %%& \citep{howiggles}. 
These extra Dirac points are found as the zero-energy solutions of 
the dispersion relation in Eq.~(\ref{eq1b_5}) for zero energy \citep{barb4}. 
%%R
% Alternatively, one can 
% study the trace of the transfer matrix, $\text{Tr}\mathcal{M}$, for energy 
% equal to zero as recently done by 
% \citep{breywiggles2}. 
%

In order to find the location of the Dirac points 
we assume $k_x = 0$, $\ve = 0$, $\mu_b = \mu_w = 0$, and consider the special 
case of 
%%R
$W_b = W_w = 1/2$ in Eq.~(\ref{eq1b_5}). 
The resulting equation
\begin{equation}\label{eq1c_8}
	1 = \cos^2 \lambda/2 + \big[(u^2/4 + k_y^2)/(u^2/4 - k_y^2)\big] \,\sin^2 \lambda/2,
\end{equation} 
has solutions for $u^2/4 - k_y^2 = u^2/4 + k_y^2$ or $\sin^2 \lambda/2 = 0$. 
This determines the values of $k_y = 0$ (at the Dirac points) as 
\begin{equation}\label{eq1c_9} 
	k_{y,j\pm} = \pm\sqrt{
	\frac{u^2}{ 4} - 4 j^2 \pi^2};
	%%R
	%%R= \pm\sqrt{ 
	%%R\Big(\frac{V_0}{2 \hbar v_F}\Big)^2 
	%%R- 	\Big(\frac{2 j \pi}{L}\Big)^2};  
\end{equation}
the extra Dirac points are for $j\neq 0$. 
%%R
%%Rand we reinserted the dimensions after 
%%Rthe second equality sign. 
For a SL spectrum symmetric around zero energy, 
the extra Dirac points are at $\ve = 0$. 
We expect from the
considerations of Sec.~\ref{sec1}(\ref{sec1b}) (and Fig. \ref{fig1_4}(b)) that for unequal 
barrier and well widths this will no longer be true. 
Indeed, in such a case the extra Dirac points shift in energy, as seen in Fig. 
\ref{fig1_4}(d), and their 
position in the spectrum is given, for $k_x = 0$, by \citep{barb4}
\begin{equation}\label{eq1c_10}
\begin{aligned}
	& \ve_{j,m} = \frac{u}{2}(1-2 W_b) + \frac{\pi^2}{2 u}\left( \frac{j^2}{W_w^2}-\frac{(j+2 m)^2}{W_b^2}\right),\\
	& {k_y}_{j,m} = \pm  
	\Big[(\ve_{j,m} + u W_b)^2 -(j \pi / W_w)^2\Big]^{1/2},
\end{aligned}
\end{equation}
where $j$ and $m$ are integers, and $m \neq 0$ corresponds to higher and lower 
crossing points. 
Also, perturbing the potential with an asymmetric term, as done by 
\citep{parkwiggles}, leads to qualitatively similar results.

\begin{figure}[ht]
  \begin{center}
	\includegraphics[width=6.5cm]{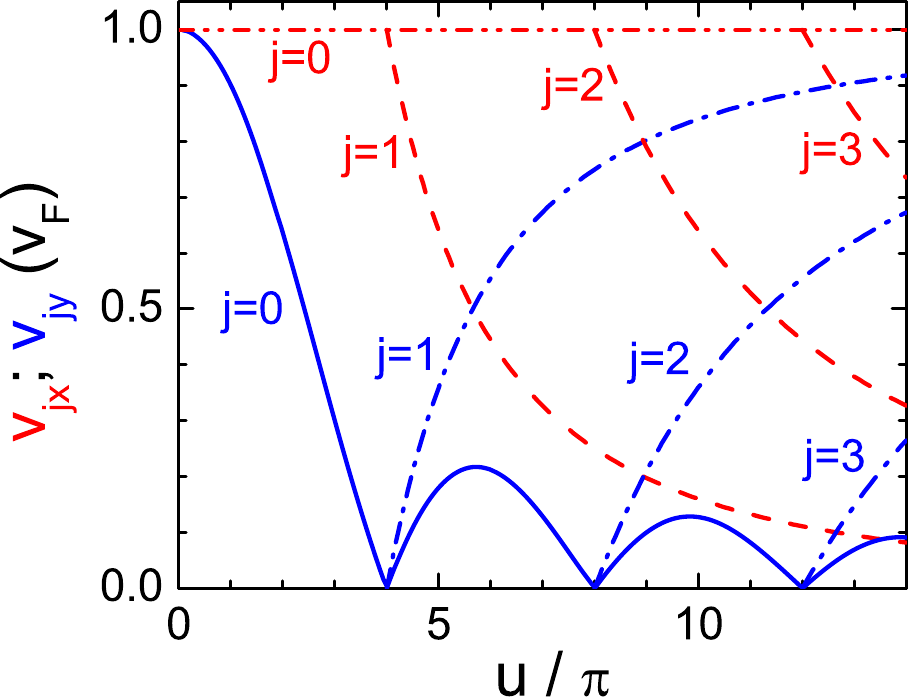}
    \end{center}
  \caption{The group velocity components 
  %%R
  $v_y$ and $v_x$ 
  at the Dirac point 
  		$j = 0$ (shown, respectively, by the solid blue and the dot-dot-dashed 
  		red curve), and at the extra Dirac points $j = 1, 2, 3$ (shown, 
  		respectively, by the dot-dashed blue and the dashed red curves) 
  		as a function of the barrier 
  		%%R
  		parameter $u = V_0 L / \hbar v_F$.
  }\label{fig1_7}
\end{figure}

An investigation of the group velocity near the (extra) Dirac points is 
appropriate for understanding the transport of carriers in the energy bands 
close to zero energy.
Near the extra Dirac points the group velocity tends to 
renormalise differently as compared to the original Dirac point. 
Near them ${\bf v}$ is oriented 
along the $y$ direction, while near the latter one ${\bf v}$
is oriented
along the $x$ direction \citep{howiggles}. 
The group velocity near the extra Dirac points can be calculated 
from Eq.~(\ref{eq1c_6}). At the $j$th extra Dirac point the magnitude of the 
velocity 
${\bf v}/v_F = (\partial \ve / \partial k_x,\,\partial \ve / \partial k_y)$ 
is given by 
\begin{equation}
\begin{aligned}
v_x/v_F &= 16 \pi^2 j^2 
%%R
\cos(k_x/2)/u^2\\
v_y/v_F &= (u^2/4 - 4 j^2 \pi^2)/u^2, 
\end{aligned}
\end{equation}
while at the main Dirac point 
%%R 
it is given by 
$v_x/v_F = 1$ and $v_y/v_F = 4\sin(u/4)/u$. 
The dependence of the velocity components on the strength 
of the potential 
barriers is shown in Fig. \ref{fig1_7}. 
From this figure we observe that new extra Dirac points emerge upon 
increasing 
%%R
$u = V_0 L / \hbar v_F$ 
(consistent with Eq.~(\ref{eq1c_9})) and $v_x$ decreases 
while $v_y$ increases. 
The Dirac point itself, however, shows a different behaviour upon increasing 
%%R
$u$, 
namely $v_x = v_F$ constant, and $v_y$ is here a globally decaying 
function showing $v_y = 0$ for periodic values of $u$, 
%%R
$u = 4 n \pi$, with $n$ 
a nonzero positive integer. 

{\it Conductivity. }  We now turn to the transport properties of a SL and look 
at the influence of 
these extra Dirac points on the conductivity. 
The diffusive dc conductivity $\sigma_{\mu \nu}$ for the SL system can be 
readily calculated from the spectrum if we assume a nearly constant relaxation 
time $\tau(E_F)\equiv \tau_F$. It is given by \citep{takis1}
\begin{equation}\label{eq_cond}
	\sigma_{\mu \nu}(E_F)=  
	\frac{e^2 \beta \tau_F}{A} 
	\sum_{n, \bf{k}}  
	v_{n\mu} v_{n\nu} f_{n\bf{k}}(1-f_{n\bf{k}}), 
\end{equation}
with $A$ the area of the system, $n$ the energy band index, $\mu,\,\nu = x,y,$ 
and $f_{n\bf{k}} = 1/[\exp(\beta(E_F - E_{n\bf{k}})) + 1]$ the 
equilibrium Fermi-Dirac distribution function; $\beta = 1 /k_B T$ and the 
temperature enters the results through 
%%R
the dimensionless value for $\beta$ which is $\beta = \hbar v_F / k_B T L = 20$. 

\begin{figure}[ht]
  \begin{center}
	\includegraphics[width=10cm]{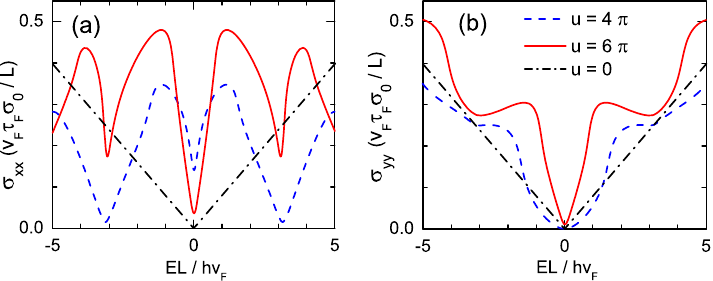} 
  \end{center}
  \caption{(Color online) 
   Conductivities, $\sigma_{xx}$ in (a) and $\sigma_{yy}$ in (b), vs Fermi 
   energy for a SL on single-layer graphene with 
   %%R
   $u = 4\pi$ and $6 \pi$ 
   for, respectively, the blue dashed and red solid curves. 
   In both cases 
   %%R
   $W_b = W_w = 0.5$. The dash-dotted black curves show the 
   conductivities in the absence of the SL potential, 
   $\sigma_{xx} = \sigma_{yy} = \ve_F \sigma_0 / 4 \pi$. 
  }\label{fig1_8}
\end{figure}

For comparison we first look at the conductivity tensor at 
zero temperature
and in the absence of a SL. 
For single-layer graphene the conductivity is given by
\begin{equation}\label{eq1c_11}
\sigma_{\mu \mu}(\ve_F)/\sigma_0 = \ve_F/4\pi
\end{equation}
with $\sigma_0 = e^2 / \hbar$, 

In Figs.~\ref{fig1_8}(a), (b) the conductivities $\sigma_{xx}$ 
and $\sigma_{yy}$ are shown for a SL as  functions of the energy.  
Notice that  for small energies the slope 
of the conductivity $\sigma_{yy}$ 
is tunable to a large extent by altering the parameter 
%%R
$u$ of the SL. 
The dashed blue curves correspond to 
%%R
$u = 4\pi$ and the rather flat 
dispersion in the $y$ direction for the lowest conduction band 
(see Fig.~\ref{fig1_5}(c,d)) 
translates to a small $\sigma_{yy}$ (for energies $EL/\hbar v_F < 1$) compared 
to the conductivity in the absence of a SL.
The solid red curves on the other hand correspond to 
%%R
$u = 6\pi$ 
and due to the extra Dirac points, which have a rather flat dispersion in the 
$x$ direction \citep{howiggles}, the conductivity $\sigma_{yy}$ is large.

\subsection{Dirac lines}\label{sec1d}
\begin{figure}[ht]
  \begin{center}
	\includegraphics[width=8cm]{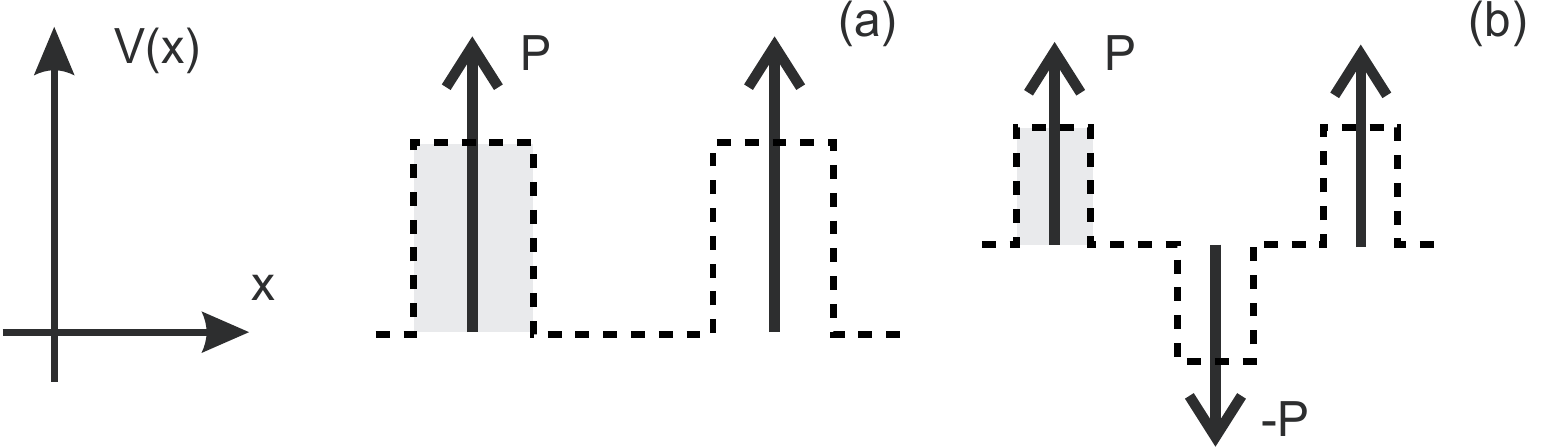}
  \end{center}
  \caption{(a) Schematics of Kronig-Penney SL on single-layer graphene. 
  (b) Extended Kronig-Penney SL.
  }\label{fig1_9}
\end{figure}
\begin{figure}[ht]
  \begin{center}
	\includegraphics[width=10.0cm]{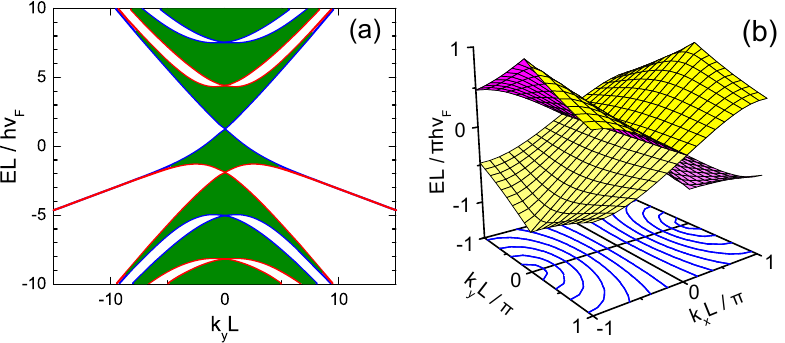}
    \end{center}
  \caption{(a) Spectrum for a Kronig-Penney SL with $P = 0.4 \pi$. 
	The blue and red curves show, respectively,  the $k_x = 0$ and $k_x = \pi/L$ 
	results which 
	delimit the energy bands (green coloured regions). 
	(b) Spectrum for an extended Kronig-Penney SL with $P = \pi/2$. 
	Notice that the Dirac point has become a Dirac line.
  }\label{fig1_10}
\end{figure}

In an effort to simplify the expressions for the dispersion relation we replace, 
as we did for the few-barrier structures, the SL barriers by $\delta$-function 
barriers. 
The square SL potential is then approximated by
\begin{equation}\label{eq1d_1}
	V(x) = P \sum_{j = - \infty}^\infty \delta(x - j L).
\end{equation}
This potential leads to the dispersion relation
\begin{equation}\label{eq1d_2}
    \cos k_x  =   \cos \lambda \cos P + (\ve/\lambda) \sin \lambda \sin P,
\end{equation}
which is {\it periodic} in $P$. This is in sharp contrast with that for standard 
electrons which is {\it not periodic} in $P$ and which in our notation reads
\begin{equation}
    \cos k_x  = \cos \lambda' + (\mu P/\lambda') \sin \lambda',
\end{equation}
where $\mu = m v_F L / \hbar$ and $\lambda'= [2\mu \ve - k_y^2]^{1/2}$. 
As can be seen from Fig.~\ref{fig1_10}(a), 
the energy band near the Dirac point has the interesting property that it 
becomes nearly flat in $k_x$, forming a plane, for large $k_y$. 
The angle which the asymptotic plane makes 
with the zero-energy plane depends on $P$ and the group velocity $v_y$ 
corresponding to this asymptotic plane  varies from $-v_F$ to $v_F$ in each 
period $n \pi < P < (n+1)\pi$. Notice that no extra Dirac points are found 
and the reason is the same as that for the asymmetric SL 
potential, i.e., the extra Dirac points shift away from zero energy. 
Alternatively, we can try to shed some light by comparing with 
Sec.~\ref{sec1}(\ref{sec1b}), where it is explained that the bound states for 
a single unit of the SL potential are similar to those of the combined single 
barrier and well.
In the region where the bound states cross 
(denoted by I in Fig.~\ref{fig1_4}(a)) 
%%R
anti-crossings 
occur and corresponding crossings in the SL spectrum (extra 
Dirac points) are expected. 
In the limit of a $\delta$-function barrier this region is reduced to a line 
(the dark green line in Fig.~\ref{fig1_4}(a)). This prevents anti-crossings 
from occurring and in this way no extra Dirac points are expected. 

{\it Extended Kronig-Penney model.}
To re-establish the symmetry between electrons and holes, as in the case of 
square barriers 
with $W_b = W_w$, we can use alternating-in-sign $\delta$-function 
barriers. The unit cell of the periodic potential contains one such barrier up, 
at $x = 0$, followed by a barrier down, at $x = L/2$,
see Fig. \ref{fig1_9}(b). The potential is given by 
\begin{equation}\label{eq1d_4}
	V(x) = P \sum_{j = - \infty}^\infty [\delta(x - j L) - \delta(x - j L - L/2)],
\end{equation}
and is the asymptotic limit of the potential shown in Fig.~\ref{fig1_1}(b). 
The resulting transfer matrix leads to the dispersion relation
\begin{equation}\label{eq1d_5}
    \cos k_x = \cos \lambda - (2 k_y^2/\lambda^2) \sin^2 (\lambda/2) \sin^2 P.
\end{equation}
This dispersion relation is {\it periodic} in $P$. As 
shown in Fig. \ref{fig1_10}(b) no extra Dirac points occur, but for the 
particular 
case of $P =(n+ 1/2)\pi$, $n$ an integer, 
the spectrum shows an interesting feature: for all $k_y$ 
we see that Eq.~(\ref{eq1d_5}) has a solution with $\ve = k_x = 0$, which 
means the Dirac point at $k_x = k_y = 0$ turned into a Dirac line along the 
$k_y$ axis. 
If we take $k_y$ not too large (of the order of $k_x$),
this spectrum has a wedge structure as was also found for rectangular SLs. 
For $k_y \rightarrow \infty$, though, the spectrum becomes a horizontal plane 
situated at $\ve = 0$. 
We can generalize this model by taking the distance $W$ between the two barriers 
of the unit cell not equal to $L/2$. This was done by 
%%R
(Ramizani Masir {\it et al.}, 2010, unpublished work). 
They found an approximate analytic expression for the 
dispersion given by
\begin{equation}\label{eq1d_6}
    \ve \approx [k_x^2 + F k_y^2]^{1/2}, \qquad F = W^2 + (L - W)^2 + 2 W(L - W)\cos(2 P).
\end{equation}
This dispersion has the shape of an anisotropic cone with a renormalized 
velocity in the $y$ direction. Comparing with Eqs. (\ref{eq1c_4}) and 
(\ref{eq1c_7}), we observe that the condition for collimation and the velocity 
renormalization in the $y$ direction is quite different for square barriers. 
For instance, in the extended KP model, with $W = L/2$, 
we find $v_y/v_F = |\cos P|$ 
while for square barriers the result is $v_y/v_F = \sin(u/4)/(u/4)$. The latter means that if 
we consider $P \equiv u/4$, the velocity in the $y$ direction is maximum 
$v_y = v_F$ for $P = (1/2 + n) \pi$ in the extended KP model while for square 
barriers $v_y = 0$ at these points.
%
%%%%%%%%%%%%%%%%%%%%%%%%%%%%%%%%%%%%%%%%%%%%%%%%%%%%%%%%%%%%%%%%%%%%%%%%%%%%%
%%%%%
%%%%%
%%%%%%%%%%%%%%%%%%%%%%%%%%%%%%%%%%%%%%%%%%%%%%%%%%%%%%%%%%%%%%%%%%%%%%%%%%%%%
%%%%%
%%%%% SECTION 2
%%%%%
%%%%%%%%%%%%%%%%%%%%%%%%%%%%%%%%%%%%%%%%%%%%%%%%%%%%%%%%%%%%%%%%%%%%%%%%%%%%%
%%%%%
%%%%%
%%%%%%%%%%%%%%%%%%%%%%%%%%%%%%%%%%%%%%%%%%%%%%%%%%%%%%%%%%%%%%%%%%%%%%%%%%%%%
%
\section{Bilayer graphene}\label{sec2}
We now turn to bilayer graphene and use again the nearest-neighbour, 
tight-binding Hamiltonian in the continuum approximation 
with ${\bf k}$ close to 
the $K$ point. If we include a potential difference between the 
two layers, the Hamiltonian is given by
\begin{equation}\label{eq2a_1}
	\mathcal{H} =  \begin{pmatrix}  U_1 & v_F\pi & t_\perp & 0 \\ v_F\pi^\dagger & U_1 & 0 & 0 \\ t_\perp & 0 & U_2 & v_F\pi^\dagger \\ 0 & 0 & v_F\pi & U_2 \end{pmatrix}.
\end{equation}
Here $U_1$ and $U_2$ are the potentials on layers
$1$ and $2$, respectively, $2\Delta=U_1-U_2$ is the potential difference, and 
$t_\perp$ describes the 
coupling between the layers. The energy spectrum for free electrons is given 
by \citep{mccann,barb2}
\begin{equation}\label{eq2a_2}
\begin{aligned}
\ve &= u_0 \pm\Big[\Delta^2 + k^2 + \frac{t^2_\perp}{2} 
+(4 \Delta^2 k^2 + k^2 t^2_\perp + \frac{t^2_\perp}{4})^{1/2}\Big]^{1/2},\\
\ve &= u_0 \pm\Big[\Delta^2 + k^2 + \frac{t^2_\perp}{2} -(4 \Delta^2 k^2 + k^2t^2_\perp + \frac{t^2_\perp}{4})^{1/2}\Big]^{1/2},
\end{aligned}
\end{equation}
with $u_{1} = u_0 + \Delta$ and $u_{2} = u_0 - \Delta$. 
%%R
Contrary to Sec. \ref{sec2} we use units in 
inverse distance, namely, 
$\ve = E/\hbar v_F$, $u_j = U_j/\hbar v_F$, and 
$k = [\lambda^2 + k_y^2]^{1/2}$. 
This spectrum exhibits an energy gap that for $2\Delta \ll t_\perp$ 
equals the difference $2\Delta$ between the conduction and 
valence band at the K point \citep{mccann}. 

Solutions for this Hamiltonian are four-vectors $\psi$ and for 1D 
potentials we can write $\psi(x,y) = \psi(x)\exp(i k_y y)$. If the potentials 
$U_{1}$ and $U_{2}$ do not vary in space, 
these solutions are of the form
\begin{equation}\label{eq2a_3}
	\Psi_\pm(x) = \kvecc{1}{f_\pm}{h_\pm}{g_\pm h_\pm}
     e^{\pm \ri \lambda x + \ri k_y y},
\end{equation}
with $f_\pm = [-\ri k_y \pm \lambda]/[\ve'-\delta]$,
$h_\pm = [(\ve'-\delta)^2 - k_y^2 - \lambda^2]/[t_\perp(\ve'-\delta)]$, and 
$g_\pm = [i k_y \pm \lambda]/[\ve'+\delta]$; 
the wave vector $\lambda$ 
is given by 
\begin{equation}\label{eq2a_4}
\lambda_\pm = \left[ \ve'^2 + \delta^2 - {k_y}^2 \pm \sqrt{ 4 \ve'^2 \delta^2 + t_\perp^2 
(\ve'^2 - \delta^2)} \right]^{1/2}.
\end{equation}
We will write $\lambda_+ = \alpha$ and $\lambda_- = \beta$.

\subsection{Tuning of the band offsets}\label{sec2b}
\begin{figure}[ht]
	\begin{center}
  	\includegraphics[width=7cm]{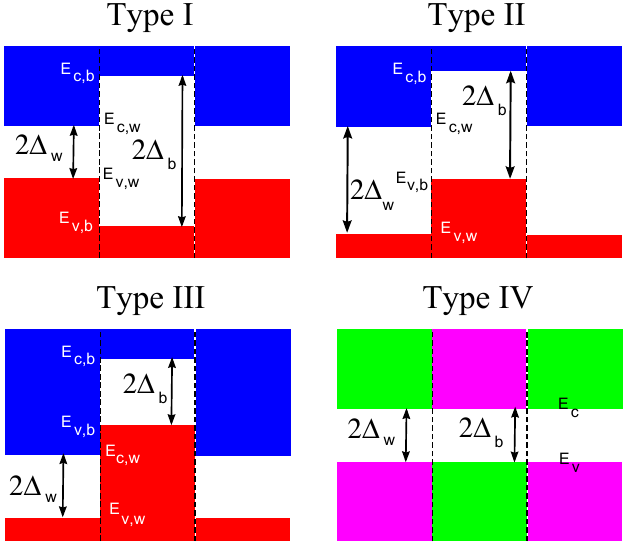}
	\end{center}
	\caption{Four different types of band alignments in bilayer graphene. 
    $E_{c,b}, E_{c,w}, E_{v,c}$, and $E_{v,b}$ denote the energies of the 
    conduction (c) and valence (v) bands in the barrier (b) and well (w) regions. The 
    corresponding gap is, respectively, $2\Delta_b$ and $2\Delta_w$. 
	}\label{fig2_2}
\end{figure}

It was shown before that using a 
1D biasing, indicated 
in Figs. \ref{fig2_2}(a,b,c) by $2\Delta$, 
one can create three types of heterostructures in graphene \citep{dragoman1}. 
A fourth type, where the energy gap 
is spatially kept constant but the bias periodically changes sign along the 
interfaces, can be introduced (see Fig.~\ref{fig2_2}(d)). We characterize 
these heterostructures as follows:\\
1) {\bf Type I:} The gate bias applied in the barrier regions is larger than in 
the well regions.\\
2) {\bf Type II:} The gaps, not necessarily equal, are shifted in energy but 
they have an overlap as shown.\\
3) {\bf Type III:} The gaps, not necessarily equal, are shifted in energy and 
have no overlap. \\ 
4) {\bf Type IV:} The bias changes sign between successive barriers and wells 
but its magnitude remains constant.\\

Type IV structures 
have been shown to localize the wave function at the interfaces 
\citep{mart,jalil}. 
To understand the influence of such interfaces in this 
section we will separately investigate structures with 
such a single interface embedded by an anti-symmetric potential. 

To describe the transmission and bound states of some simple structures 
we notice that in the energy region of interest, i.e., for 
$|E| < t_\perp$, the eigenstates which are propagating are the ones with 
$ \lambda = \alpha$. Accordingly, 
from now on we will assume that $\beta$ is 
complex. In this way we can simply use the transfer-matrix approach of 
Sec.~\ref{sec1} in the transmission calculations. This leads 
to the relation 
\begin{equation}\label{eq2b_1}
    \kvecc{t}{0}{e_d}{0} = \mathcal{N} \kvecc{1}{r}{0}{e_g}.
\end{equation}
Again the transmission is given by $T = |t|^2$. 

\begin{figure}[ht]
	\begin{center}
  	\includegraphics[width=11cm]{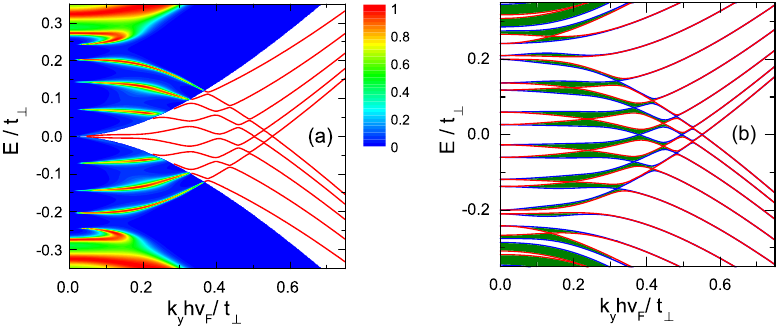}
	\end{center}
	\caption{%
	(a) Contour plot of the transmission for the potential of 
	Fig.~\ref{fig1_1}(b) in 
	bilayer graphene with $W_b = W_w = 40$ nm, $V_b = -V_w = 100$ meV 
	and zero bias. Bound states are shown by the red curves. 
	(b) Spectrum 
	for a SL whose unit is the potential structure of Fig.~\ref{fig1_1}(b). 
	Blue and red curves show, respectively, the $k_x = 0$ and $k_x = \pi/L$ 
	results which 
	delimit the energy bands (green coloured regions). 
	}\label{fig2_3}
\end{figure}

For a single barrier the transmission in bilayer graphene is 
given by a 
complicated expression. Therefore, we will first look at a few limiting cases.  
First we assume a zero bias $\Delta = 0$ that corresponds to 
a particular case of 
type III heterostructures. 
In this case we slightly change the 
definition of the wave vectors: 
for $\Delta = 0$ we assume
%%& 
$\alpha (\beta) = [\ve^2 +(-)\ve t_\perp - k_y^2]^{1/2}$. 
If we restrict the motion along the 
$x$ axis, by taking $k_y = 0$, and assume a bias $\Delta = 0$, then the 
transmission $T = |t|^2$ is given via
\begin{equation}
\begin{aligned}
	1/t &= e^{i \alpha_0 D}[\cos (\alpha_b D) - i Q \sin (\alpha_b D)],\\
	Q &= \frac{1}{2}\left(\frac{\alpha_b \ve_0}{\alpha_0 \ve_b} + \frac{\alpha_0 \ve_b}{\alpha_b \ve_0} \right).
\end{aligned}
\end{equation}
This expression  depends only on the propagating wave vector $\alpha$ 
($\beta$ for $E < 0$) as 
propagating and localized states are decoupled in this approximation. 
This  also means that one does not find any resonances in the transmission for 
energies in the barrier region, i.e., for \ $0 < \ve < u$. Due to the 
coupling for nonzero $k_y$ with the 
localized states, resonances in the transmission will occur 
(see Fig.~\ref{fig2_3}). We can easily generalize 
this expression to account for the double 
barrier case under the same assumptions. With an inter-barrier distance $W_w$ 
one obtains the transmission \citep{barb2} 
$T_d = |t_d|^2$ from 
\begin{equation}\label{eqtd}
    t_d = \frac{e^{i 2 \alpha_0 (W_w + 2 W_b)}|t|^2 e^{i 2 \phi_t}}{1-|r|^2 e^{i 2 \phi_r} e^{i 2 \alpha_0 W_w}},
\end{equation}
with $r = |r|\re^{\ri\phi_r}$, and $t = |t|\re^{\ri\phi_t}$, corresponding to 
the single barrier transmission and reflection amplitudes.
In this case we do have 
resonances due to the well states; they occur 
for $\re^{\ri 2 \phi_r} \re^{\ri 2 \alpha_0 W_w} = 1$. As $\phi_r$ is independent of 
$W_w$, one obtains more resonances by increasing $W_w$.

For a single $\delta$-function barrier with 
potential $V(x)/\hbar v_F = P \delta(x)$ under zero bias, we find 
the transmission amplitude
\begin{equation}
	1/t = \cos P  + \ri\mu \sin P
	 + \frac{(\alpha - \beta)^2 k_y^2}{4 \alpha \beta \ve^2} \frac{\sin P}{\cot P + \ri\nu},	 
\end{equation}
where $\mu=(\ve + 1/2)/\alpha$ and $\nu=(\ve - 1/2)/\beta$. 
Notice that this formula is {\it periodic} in the strength of the 
barrier $P$ as in the single-layer case.

\begin{figure}[tb]
	\begin{center}
	\includegraphics[width=10cm]{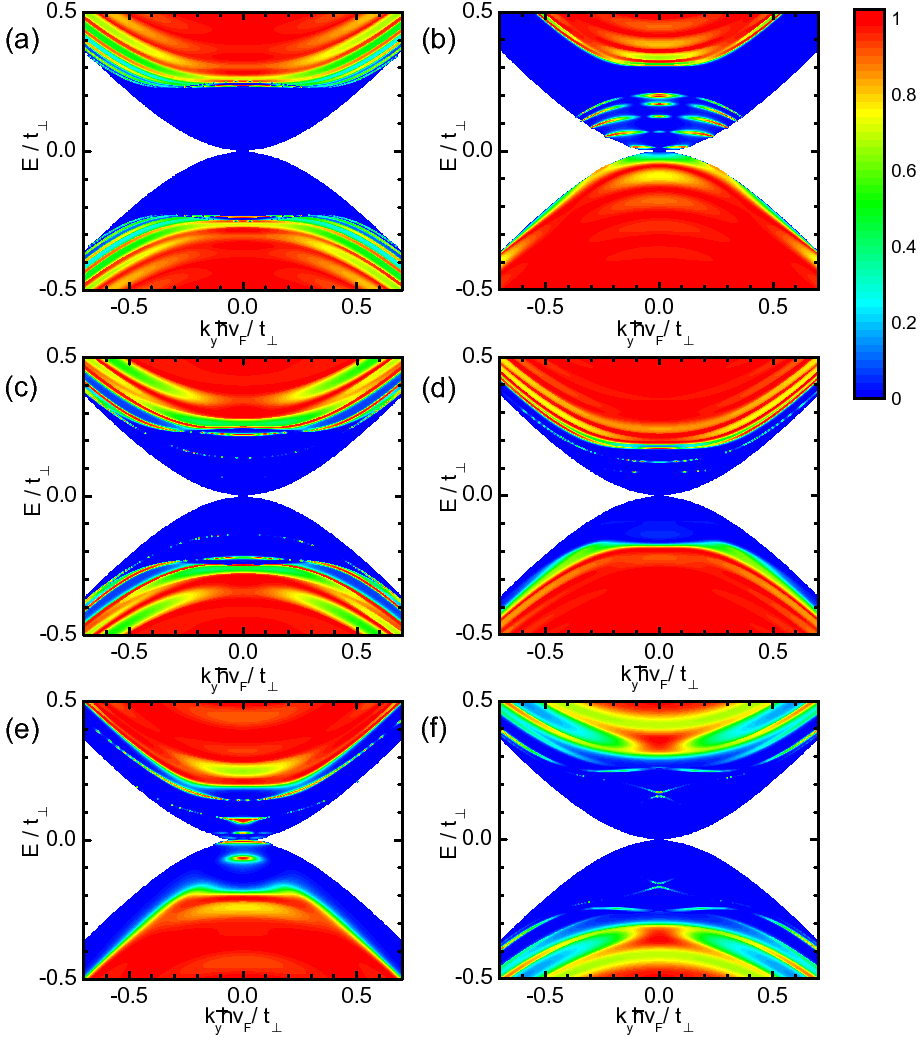}
	\end{center}
	\caption{\label{fig2_5} 
	(Color online) Contour plot of the transmission through a single barrier in
    (a) and (b),  for width $W_b = 50$ nm, and through double barriers in (c), (d), 
    (e), and (f) of equal widths $W_b = 20$ nm that are separated by $W_w = 20$ 
    nm. 
    Other parameters are as follows:
    (a) $\Delta_b = 100$ meV, $V_b = 0$ meV. 
    (b) $\Delta_b = 20$ meV, $V_b = 50$ meV.
    (c) \emph{Type I}: $V_b = V_w = 0$ meV, $\Delta_w = 20$ meV, and 
    $\Delta_b = 100$ meV. 
    (d) \emph{Type II}: $Vb = -Vw = 20$ meV, $\Delta_w = \Delta = 50$ meV, 
    (e) \emph{Type III}: $V_b = - V_w = 50$ meV, $\Delta_w = \Delta_b = 20$ meV. 
    (f) \emph{Type IV}: $V_b = V_w = 0$ meV, $\Delta_b = -\Delta_w = 100$ meV.}
\end{figure}
\begin{figure}[tpb]
	\begin{center}
	\includegraphics[width=7cm]{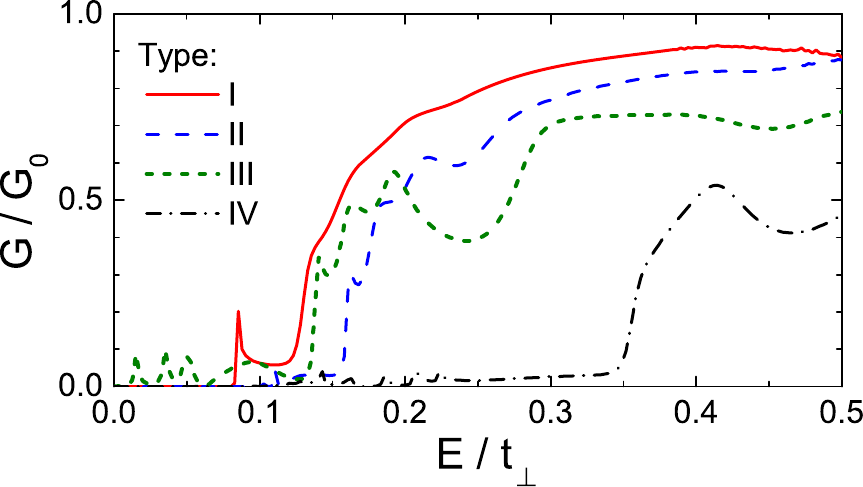}
	\end{center}
    \caption{\label{fig2_5G}
    (Color online) Two-terminal conductance of four equally spaced barriers vs energy for 
    $W_b = W_w = 10$ nm and different SL types I-IV. 
    The solid red curve (type I) is for $\Delta_b = 50$ meV, $\Delta_w = 20$ 
    meV, and $V_w = V_b = 0$. 
    The blue dashed curve (type II) is for $\Delta_b = \Delta_w = 50$ meV and
    $V_b = -V_w = 20$ meV. 
    The green dotted curve (type III) is for 
    $\Delta_b = \Delta_w = 20$ meV and $V_b = - V_w = 50$ meV. 
    The black dash-dotted curve (type IV) is for $\Delta_b = -\Delta_w = 50$ 
    meV and $V_w = V_b = 0$.}
\end{figure}
For the general case we obtained numerical results for the 
transmission 
through various types of single and double 
barrier structures;  they are shown in Fig.~\ref{fig2_5}. 
The different types of structures clearly lead to different behaviour of the 
tunnelling resonances. 

An interesting structure to study is the fourth type of SLs 
shown in Fig.~\ref{fig2_2}(d). 
To investigate the influence of the localized 
states \citep{mart,jalil} on the transport properties we embed the 
anti-symmetric potential profile in a structure with unbiased layers. 
\begin{figure}[tbh]
	\begin{center}
	\includegraphics[width=10cm]{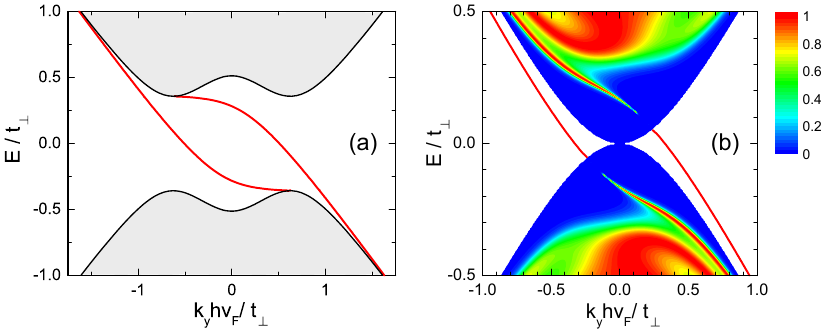}
	\end{center}
	\caption{%
	(a) Bound states of the anti-symmetric potential profile (type IV) with 
	 bias $\Delta_w = -\Delta_b = 200$ meV. (b) Contour plot of the transmission 
	through a $20$ nm wide barrier 
	consisting of two regions with opposite biases 
	$\Delta = \pm100$ meV.
	}\label{fig2_4}
\end{figure}

{\it Conductance} At zero temperature $G$ can be calculated 
from the transmission using Eq. (13) 
with $G_0=(4 e^2 L_y /2 \pi h)\,(E_F^2+t_\perp E_F)^{1/2}/\hbar v_F$ for bilayer 
graphene and $L_y$ the width of the sample. 
The angle of incidence $\phi$ is given by $\tan \phi = k_y/\alpha$ 
with $\alpha$ the wave vector  outside the barrier. Figure \ref{fig2_5G} shows $G$ 
for the four SL types. Notice the  clear differences in 
1) the onset of the conductance and  2) the number and amplitude of the oscillations.

{\it Bound states.} To describe 
bound states we assume that there are no 
propagating states, i.e., $\alpha$ and $\beta$ are imaginary or complex (the 
latter case can be solved separately), and only the eigenstates 
with exponentially decaying behaviour are nonzero leading to the relation 
\begin{equation}\label{eq2b_3}
    \kvecc{f_d}{0}{e_d}{0} = \mathcal{N} \kvecc{0}{f_g}{0}{e_g}.
\end{equation}
From this relation we can find the dispersion relation for the bound states.

To study the localized states for the anti-symmetric potential 
profile \citep{mart,jalil} we will use a sharp kink 
profile (step function). 
The spectrum found by the method above is shown in Fig.~\ref{fig2_4}(a). We see 
that 
there are two bound states, both with negative group velocity 
$v_y \propto \partial{\ve}/\partial{k_y}$, as found previously 
by 
\citep{mart}. No bound state near zero energy was found for 
$k_y \rightarrow \infty$ in contradiction with 
\citep{jalil}. 
For zero energy we find the solution 
\begin{equation}
\begin{aligned}
	k_y &= \pm\frac{1}{2}
	[\Delta^2 + (\Delta^4 + 2 \Delta^2 t_\perp^2)^{1/2}]^{1/2}\\
	&\approx \pm \sqrt{\Delta t_\perp}/2^{3/4}, \quad 
	\Delta \ll t_\perp;
\end{aligned}
\end{equation}
the approximation on the second line leads to the expression found 
by \citep{mart}.
\subsection{Superlattices}\label{sec2c}
The heterostructures above (see Fig. \ref{fig2_2}), can be used to create 
four different types of SLs \citep{dragoman1}. We will especially focus on type IV 
and type III SLs in certain limiting cases.

\begin{figure}[tbh]
	\begin{center}
	\includegraphics[width=13cm]{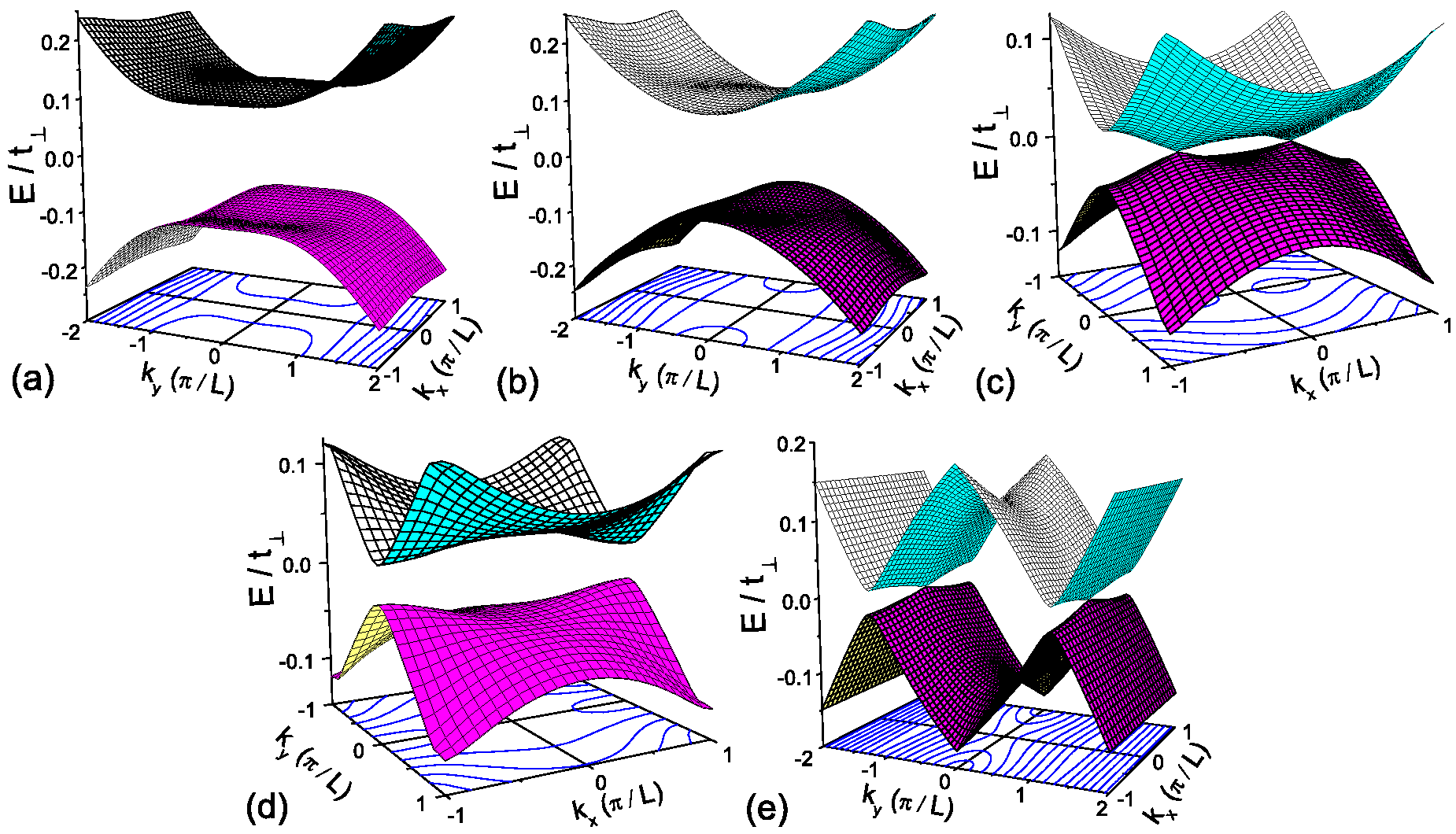}
	\end{center}
    \caption{\label{fig2_6} 
    (Color online) Lowest conduction and highest valence band of the 
    spectrum for a square SL with period $L = 20$ nm and $W_b = W_w = 10$ nm. 
	(a) \emph{Type I}: $\Delta_b = 100$ meV and $\Delta_w = 0$. 
	(b) \emph{Type II}: As in (a) for $\Delta_b = \Delta_w = 50$ meV, and 
		$V_b = -V_w = 25$ meV. 
	(c) \emph{Type III}: $V_b = - V_w = 25$ meV, 
	and $\Delta_b = \Delta_w = 0$. 
	(d) \emph{Type III}: $V_b = - V_w = 50$ meV  and $\Delta_b = \Delta_w = 0$. 
	(e) \emph{Type IV}: Plot of the spectrum for a square SL with average 
    	potential $V_b = V_w = 0$ and $\Delta_b = - \Delta_w = 100$ meV. 
    	The contours are for the conduction band and show that the dispersion is 
    	almost flat in the $x$ direction.}
\end{figure}
%Type I
For a  type I SL we see in Fig.~\ref{fig2_6}(a) that the conduction and valence 
band of the bilayer structure are qualitatively similar to those 
in the presence of a uniform bias. 
%Type II
Type II structures maintain this gap, see 
Fig. \ref{fig2_6}(b), as there is a range in energy for which there is a gap in 
the  SL potential in the barrier and well regions. 
%Type III
In type III structures we have two interesting features, which can close the 
gap. 
First we see from Fig.~\ref{fig2_3}(b) that for zero bias, similar to 
single-layer graphene, extra Dirac 
points appear for $k_x = 0$, likewise for Fig.~\ref{fig1_4}(d). 
%%R
In the case $W_b = W_w = L/2 = W$, $k_x = 0$ 
and $E = 0$ the values for the $k_{y}$ where extra Dirac points occur are given 
by the following transcendental equation
\begin{equation}
	[\cos(\alpha W) \cos(\beta W) - 1] + \frac{\alpha^2 + \beta^2 - 4 ky^2}{2\alpha \beta} 
	\sin(\alpha W) \sin(\beta W) = 0.
\end{equation}
Comparing the figures 
%%R Comparing these last two figures 
\ref{fig2_3}(b) and \ref{fig1_4}(d) 
we remark that, different from the single-layer case, 
for bilayer graphene the bands in the barrier region are not only flat in the 
$x$ direction for large $k_y$ values but also for 
small $k_y$. The latter corresponds to the zero transmission value inside the 
barrier region for tunneling through a single unbiased barrier in bilayer 
graphene. 
Secondly, if there are no extra Dirac points (small parameter $u L$) for 
certain SL parameters, the gap closes at two 
%%R Dirac 
points 
%%R
at the Fermi-level 
for $k_y = 0$. 
The latter we will investigate a bit more in the extended Kronig-Penney model.
%Type IV
Periodically changing the 
sign of the bias (type IV) introduces a splitting of 
the charge neutrality point along 
the $k_y$ axis; this agrees 
with what was found by \citep{mart}. We illustrate that 
in Fig.~\ref{fig2_5}(e) for a SL with $\Delta_b = -\Delta_w = 100$ meV. 
We also see  that the two valleys in the spectrum are rather flat in the $x$ 
direction. Upon increasing the parameter $\Delta L$, the two touching points 
shift to larger $\pm k_y$ and the valleys become flatter in the $x$ direction. 
For all four types of SLs the spectrum is anisotropic and results in 
very different velocities along the $x$ and $y$ directions. 

{\it Extended Kronig-Penney model.} 
To understand which SL parameters lead to the creation of a gap we look at the 
Kronig-Penney limit of  type III SLs for zero bias (Barbier {\it et al.} 2010, 
unpublished work). 
Also we 
choose the extended 
Kronig-Penney model to ensure spectra symmetric with respect to 
the zero-energy value, 
such that the zero-energy solutions can be traced down more easily. If the 
latter zero modes exist, there is no gap. To simplify the calculations we restrict the 
spectrum to that for 
$k_y = 0$. This assumption is certainly not valid if the parameter $u L$ is large because 
in that case we expect extra Dirac points (not in the KP limit) to appear that 
will close the gap. 
The spectrum for $k_y = 0$ is determined by the 
transcendental equations
\begin{subequations}
\begin{eqnarray}\label{eq7_2}
	\cos k_x L  &=& \cos \alpha L \cos^2 P + D_\alpha \sin^2 P ,\\
	\cos k_x L & =& \cos \beta L \cos^2 P + D_\beta \sin^2 P,
\end{eqnarray}
\end{subequations}  
with $D_{\lambda} = \left[(\lambda^2 + \ve^2) \cos \lambda L - \lambda^2 + \ve^2 \right]/4 \lambda^2 \ve^2$, and $\lambda = \alpha, \, \beta$. 
To see whether 
there is a gap in the spectrum we look for a solution 
with $\ve = 0$ in the dispersion relations. This gives 
two values for $k_x$ where zero energy solutions occur
\begin{equation}\label{eq7_3}
	k_{x,0}  = \pm\arccos[ 1 -(L^2/8) \sin^2 P ]/L,
\end{equation}
and the crossing points are at $(\ve,\,k_x,\,k_y) = (0,\,\pm k_{x,0},\,0)$. If 
the $k_{x,0}$ value is not real, then there is no solution at zero energy and a 
gap arises in the spectrum. From Eq. (\ref{eq7_2}) we see that for 
$\sin^2 P > 16/L^2$ a band gap arises. 

{\it Conductivity. } 
In bilayer graphene the diffusive dc conductivity, 
given by 
%%R
Eq.~(\ref{eq_cond}), 
takes the form
\begin{equation}
\sigma_{\mu \mu}(\ve_F)/\sigma_0 =  
(k_F^3/4\pi\ve_F^2)\Big[1\pm \delta/2(k_F^2\delta + 1/4)^{1/2}\Big]^2,
\end{equation}
with $k_F = [\ve_F^2 + \Delta^2 \mp (\ve_F^2 \delta - \Delta^2)^{1/2}]^{1/2}$, 
$\delta = 1 + 4\Delta^2$, and 
$\sigma_0 = e^2 \tau_F t_\perp/\hbar^2$.

\begin{figure}[thb]
	\begin{center}
	\includegraphics[width=10cm]{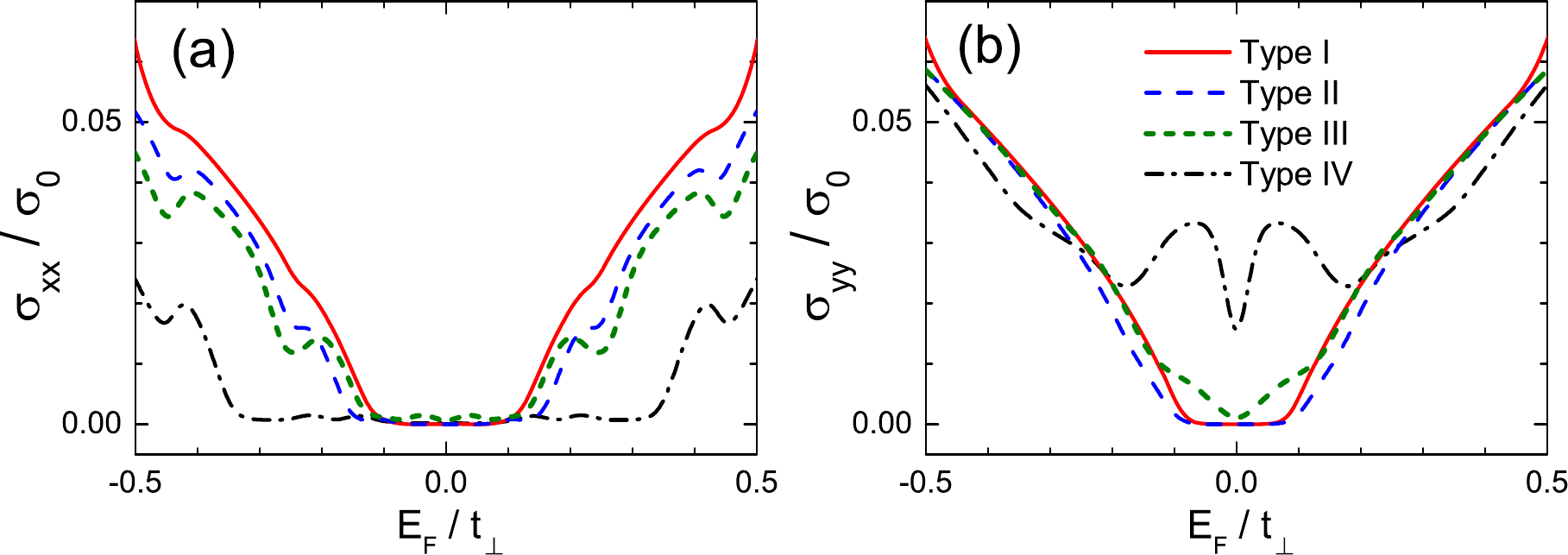}
	\end{center}
    \caption{\label{fig2_7}
    (Color online) Conductivities,  $\sigma_{xx}$ in (a) and $\sigma_{yy}$ 
   in  (b), vs Fermi energy  for the four types of SLs with 
   $L = 20$ nm and $W_b = W_w = 10$ nm, 
    at temperature $T = 45 K$;  $\sigma_0 = e^2 \tau_F t_\perp/\hbar^2$. 
    \emph{Type I}: $\Delta_b = 50$ meV, $\Delta_w = 25$ meV and $V_b = V_w = 0$. 
    \emph{Type II}: $\Delta_b = \Delta_w = 25$ meV and $V_b = -V_w = 50$ meV.
    \emph{Type III}: $\Delta_b = \Delta_w = 50$ meV and $V_b = - V_w = 25$ meV. 
    \emph{Type IV}: $\Delta_b = -\Delta_w = 100$ meV and $V_b = V_w = 0$.}
\end{figure}
In Figs.~\ref{fig2_7}(a), (b) the conductivities $\sigma_{xx}$ in (a) and $\sigma_{yy}$ in (b) 
for bilayer graphene are 
shown for the various types 
of SLs defined in Sec.~\ref{sec2}(\ref{sec2c}). 
Notice that for type IV SL the conductivities $\sigma_{xx}$ 
and $\sigma_{yy}$ differ substantially due to the anisotropy 
in the spectrum.

%%%%%%%%%%%%%%%%%%%%%%%%%%%%%%%%%%%%%%%%%%%%%%%%%%%%%%%%%%%%%%%%%%%%%%%%%%%%%
%%%%%
%%%%%
%%%%%%%%%%%%%%%%%%%%%%%%%%%%%%%%%%%%%%%%%%%%%%%%%%%%%%%%%%%%%%%%%%%%%%%%%%%%%
%%%%%
%%%%% CONCLUSION
%%%%%
%%%%%%%%%%%%%%%%%%%%%%%%%%%%%%%%%%%%%%%%%%%%%%%%%%%%%%%%%%%%%%%%%%%%%%%%%%%%%
%%%%%
%%%%%
%%%%%%%%%%%%%%%%%%%%%%%%%%%%%%%%%%%%%%%%%%%%%%%%%%%%%%%%%%%%%%%%%%%%%%%%%%%%%
\section{Conclusions}\label{sec3}
We reviewed the electronic band structure of single-layer and bilayer graphene 
in the presence of 1D periodic potentials. In addition, we investigated the 
conditions that lead to carrier collimation in single-layer graphene and 
determined when extra Dirac points appear in the spectrum and what their 
influence is on the conductivity. Furthermore, we investigated the tunnelling 
through, and bound states created by, simple barrier structures. 
In single-layer graphene we found that the SL spectrum can be linked to the 
bound states of a combined barrier and a well. 

In bilayer graphene we considered transport through different types of 
heterostructures, where we distinguished between four types of band alignments. 
We also connected  the bound states in an anti-symmetric 
potential (type IV) with the transmission through such a potential barrier. 
Furthermore, we investigated the same four types of band alignments in SLs. 
The differences between the four types of SLs are reflected not only in the 
spectrum but also in the conductivities parallel and perpendicular to the SL 
direction. For type III SLs, which have a zero bias,  we found a feature in the 
spectrum similar to the extra Dirac points found for single-layer graphene. 
Also, for not too large strengths of the SL barriers we found that the 
valence and condunction bands touch at points in {\bf k} space with $k_y = 0$ 
and nonzero $k_y$. Type IV SLs tend to split the K (K') valley into two valleys.

In the Kronig-Penney limit, where we take the barriers to be $\delta$ functions 
$V(x)/\hbar v_F = P \delta(x)$, we saw that the SL spectra, 
the transmission, conductance, etc., are periodic in the strength of 
the barriers. As s well known, this is not the case for standard electrons. 
An important qualitatively new feature is encountered in the extended 
Kronig-Penney limit for 
%%&
$P=(n+1/2)\pi$, see Sec.~\ref{sec1}(\ref{sec1d}): the Dirac point 
becomes a Dirac line.

We expect that these relatively recent findings, that we reviewed in this work, 
will be tested experimentally in the near future.
\begin{acknowledgements}
This work was supported by IMEC,
the Flemish Science Foundation (FWO-Vl), the Belgian Science Policy
(IAP), and the Canadian NSERC Grant No. OGP0121756. 
\end{acknowledgements}
%
%%%%%%%%%%%%%%%%%%%%%%%%%%%%%%%%%%%%%%%%%%%%%%%%%%%%%%%%%%%%%%%%%%%%%%%%%%%%%
%%%%%
%%%%%
%%%%%%%%%%%%%%%%%%%%%%%%%%%%%%%%%%%%%%%%%%%%%%%%%%%%%%%%%%%%%%%%%%%%%%%%%%%%%
%%%%%
%%%%% REFERENCES
%%%%%
%%%%%%%%%%%%%%%%%%%%%%%%%%%%%%%%%%%%%%%%%%%%%%%%%%%%%%%%%%%%%%%%%%%%%%%%%%%%%
%%%%%
%%%%%
%%%%%%%%%%%%%%%%%%%%%%%%%%%%%%%%%%%%%%%%%%%%%%%%%%%%%%%%%%%%%%%%%%%%%%%%%%%%%
%

%
\end{document}